\documentclass[twocolumn,trackchanges]{aastex63}

\newcommand{\teff}{$T_{\rm eff}$~}
\newcommand{\teffns}{$T_{\rm eff}$}

\newcommand{\logg}{$\log g$~}
\newcommand{\loggns}{$\log g$}

\newcommand{\spirou}{SPIRou~}
\newcommand{\spirouns}{SPIRou}

\received{September 12, 2023}
\revised{February 27, 2024}
\accepted{March 4, 2024}

\shorttitle{Spectroscopy of Barnard's star with \spirou}
\shortauthors{Jahandar et al.}
\graphicspath{{./}{figures/}}

\begin{document}

\title{Comprehensive High-resolution Chemical Spectroscopy of Barnard's star with \spirou}

\correspondingauthor{Farbod Jahandar}
\email{farbod.jahandar@umontreal.ca}

\author[0000-0003-0029-2835]{Farbod Jahandar} 
\affiliation{Trottier Institute for Research on Exoplanets, Département de Physique, Université de Montréal, 1375 Ave Thérèse-Lavoie-Roux, Montréal, QC, H2V 0B3, Canada}

\author[0000-0001-5485-4675]{Ren\'e Doyon} 
\affiliation{Trottier Institute for Research on Exoplanets, Département de Physique, Université de Montréal, 1375 Ave Thérèse-Lavoie-Roux, Montréal, QC, H2V 0B3, Canada}
\affiliation{Observatoire du Mont-M\'egantic, Universit\'e de Montr\'eal, Montr\'eal H3C 3J7, Canada}

\author[0000-0003-3506-5667]{\'Etienne Artigau} 
\affiliation{Trottier Institute for Research on Exoplanets, Département de Physique, Université de Montréal, 1375 Ave Thérèse-Lavoie-Roux, Montréal, QC, H2V 0B3, Canada}
\affiliation{Observatoire du Mont-M\'egantic, Universit\'e de Montr\'eal, Montr\'eal H3C 3J7, Canada}

\author[0000-0003-4166-4121]{Neil J. Cook} 
\affiliation{Trottier Institute for Research on Exoplanets, Département de Physique, Université de Montréal, 1375 Ave Thérèse-Lavoie-Roux, Montréal, QC, H2V 0B3, Canada}

\author[0000-0001-9291-5555]{Charles Cadieux}
\affiliation{Trottier Institute for Research on Exoplanets, Département de Physique, Université de Montréal, 1375 Ave Thérèse-Lavoie-Roux, Montréal, QC, H2V 0B3, Canada}

\author[0000-0002-6780-4252]{David Lafreni\`ere}
\affiliation{Trottier Institute for Research on Exoplanets, Département de Physique, Université de Montréal, 1375 Ave Thérèse-Lavoie-Roux, Montréal, QC, H2V 0B3, Canada}

\author[0000-0003-0536-4607]{Thierry Forveille} 
\affiliation{Univ.\ Grenoble Alpes, CNRS, IPAG, 38000 Grenoble, France}

\author[0000-0001-5541-2887]{Jean-Fran\c cois Donati} 
\affiliation{Universit\'e de Toulouse, CNRS, IRAP, 14 Avenue Belin, 31400 Toulouse, France}

\author[0000-0002-1436-7351]{Pascal Fouqu\'e}
\affiliation{Canada-France-Hawaii Telescope, CNRS, Kamuela, HI 96743, USA}
\affiliation{Universit\'e de Toulouse, CNRS, IRAP, 14 Avenue Belin, 31400 Toulouse, France}

\author[0000-0003-2471-1299]{Andr\'es Carmona} 
\affiliation{Univ.\ Grenoble Alpes, CNRS, IPAG, 38000 Grenoble, France}

\author[0000-0001-5383-9393]{Ryan Cloutier}
\affiliation{Department of Physics \& Astronomy, McMaster University, 1280 Main St W, Hamilton, ON L8S 4L8, Canada}

\author[0000-0003-4019-0630]{Paul Cristofari} 
\affiliation{Center for Astrophysics $\vert{}$ Harvard \& Smithsonian, 60 Garden Street, Cambridge, MA, 02138, USA}

\author[0000-0002-5258-6846]{Eric Gaidos} 
\affiliation{University of Hawai'i at M\={a}noa, Department of Earth Sciences, Honolulu, HI 96822, USA}

\author[0000-0001-8056-9202]{Jo\~ao Gomes da Silva} 
\affiliation{Instituto de Astrofísica e Ciências do Espaço, Universidade do Porto, CAUP, Rua das Estrelas, 4150-762, Porto, Portugal}

\author[0000-0002-8786-8499]{Lison Malo}
\affiliation{Trottier Institute for Research on Exoplanets, Département de Physique, Université de Montréal, 1375 Ave Thérèse-Lavoie-Roux, Montréal, QC, H2V 0B3, Canada}
\affiliation{Observatoire du Mont-M\'egantic, Universit\'e de Montr\'eal, Montr\'eal H3C 3J7, Canada}

\author[0000-0002-5084-168X]{Eder Martioli}
\affiliation{Laborat\'{o}rio Nacional de Astrof\'{i}sica, Rua Estados
Unidos 154, 37504-364, Itajub\'{a} - MG, Brazil}
\affiliation{Institut d'Astrophysique de Paris, CNRS, UMR 7095, Sorbonne
Universit\'{e}, 98 bis bd Arago, 75014 Paris, France}

\author[0000-0001-7804-2145]{J.-D.~do~Nascimento,~Jr.}
\affiliation{Universidade Federal do Rio Grande do Norte (UFRN), Departamento de F\'isica, 59078-970, Natal, RN, Brazil}
\affiliation{Center for Astrophysics $\vert{}$ Harvard \& Smithsonian, 60 Garden Street, Cambridge, MA, 02138, USA}

\author[0000-0002-8573-805X]{Stefan Pelletier}
\affiliation{Trottier Institute for Research on Exoplanets, Département de Physique, Université de Montréal, 1375 Ave Thérèse-Lavoie-Roux, Montréal, QC, H2V 0B3, Canada}

\author[0000-0002-5922-8267]{Thomas Vandal} 
\affiliation{Trottier Institute for Research on Exoplanets, Département de Physique, Université de Montréal, 1375 Ave Thérèse-Lavoie-Roux, Montréal, QC, H2V 0B3, Canada}

\author[0000-0003-4134-2042]{Kim Venn}
\affiliation{Department of Physics and Astronomy, University of Victoria, Victoria, BC, V8W 3P2, Canada}

\begin{abstract}

Determination of fundamental parameters of stars impacts all fields of astrophysics, from galaxy evolution to constraining the internal structure of exoplanets. This paper presents a detailed spectroscopic analysis of Barnard's star that compares an exceptionally high-quality (an average signal-to-noise ratio of $\sim$1000 in the entire domain), high-resolution NIR spectrum taken with CFHT/SPIRou to PHOENIX-ACES stellar atmosphere models. The observed spectrum shows thousands of lines not identified in the models with a similar large number of lines present in the model but not in the observed data. We also identify several other caveats such as continuum mismatch, unresolved contamination and spectral lines significantly shifted from their expected wavelengths, all of these can be a source of bias for abundance determination. Out of $>10^4$ observed lines in the NIR that could be used for chemical spectroscopy, we identify a short list of a few hundred lines that are reliable.
We present a novel method for determining the effective temperature and overall metallicity of slowly-rotating M dwarfs that uses several groups of lines  as  opposed to bulk spectral fitting methods. With this method, we infer \teffns\,=\,3231\,$\pm$\,21\,K for Barnard's star, consistent with the value of 3238\,$\pm$\,11\,K  inferred from the interferometric method. We also provide abundance measurements of 15 different elements for Barnard's star, including the abundances of four elements (K, O, Y, Th) never reported before for this star. This work emphasizes the need to improve current atmosphere models to fully exploit the NIR domain for chemical spectroscopy analysis.

\end{abstract}



\section{Introduction} \label{sec:intro}

M dwarfs are the most common type (\citealt{henry1994solar}; \citealt{winters2019solar}; \citealt{Reyle_2021}) of stars in the Galaxy. The determination of fundamental parameters of M dwarfs impacts many fields of astrophysics, from the study of the chemical evolution of stars to the internal modeling of their potential exoplanets' interior structure. The typical mass and radius of M dwarfs lie within 0.1-0.74 M$_{\odot}$ and 0.1-0.67 R$_{\odot}$ (\citealt{mann2015constrain}; \citealt{reiners2018carmenes}) making them the smallest types of stars in the main sequence. As an immediate result of their low mass and almost fully convective nature, the hydrogen-burning timescale in M dwarfs can be much longer than the age of the Universe, often several hundred Gyr depending on the mass of the star \citep{chabrier2000theory}. The small size and low luminosity of M dwarfs, result in their habitable zone being much closer to the star. Consequently, planets within this habitable zone naturally have shorter orbital periods, which makes them ideal targets for searching and characterizing HZ exoplanets via transit (e.g., \citealt{muirhead2014characterizing}; \citealt{martinez2017stellar}) and radial velocity methods (\citealt{campbell1988search}; \citealt{latham1989unseen}; \citealt{mayor1995jupiter}). Occurrence rate studies (e.g., \citealt{bonfils2013harps}; \citealt{dressing2015occurrence}; \citealt{mulders2015increase}; \citealt{gaidos2016they}; \citealt{Cloutier-Menou_2020}; \citealt{Hsu_2020}) have also shown that M dwarfs host more short-period planets than more massive stars, and the Kepler mission's observations have revealed the planet occurrence rate of 2.5\,$\pm$\,0.2 planets per M dwarf for the targets with radii 1-4 R$_{\oplus}$ and period $<$ 200 days (\citealt{dressing2015occurrence}).


Various methods have been used for determining the fundamental parameters of stars. The effective temperature (\teffns) is determined through medium- and high-resolution spectroscopy (e.g., \citealt{lamb2016using}; \citealt{rajpurohit2018exploring}), photometry (e.g., \citealt{casagrande2008m}) as well as the bolometric method involving interferometric measurements of the stellar radius (\citealt{boyajian2012stellar}). While the interferometry method is arguably the most accurate way of determining \teff independent of synthetic models, measuring angular diameters is only feasible for the nearest M dwarfs given the maximum baselines of current optical/infrared interferometers.  Photometry also has its caveats. Even with the best photometric calibrations, photometric estimates  of \teff of some M dwarfs can show up to 2$\sigma$ offset from spectroscopic values (\citealt{souto2020stellar}). These challenges emphasize the need for other complementary estimates inferred from high-resolution spectroscopy.

The spectroscopy method is also not without challenges. Due to the low surface temperature of M dwarfs (i.e., 2500\,K$<$\teffns$<$4000\,K), their spectra are generally dominated by the extensive blends of molecular bands (e.g., TiO, VO, OH, CO, H$_2$O) and absorption lines that make the detection of the continuum level of the spectrum difficult. Nevertheless, a compelling reason for conducting spectroscopy in the NIR rather than in the optical is that it is where most of the stellar flux is concentrated for M dwarfs. 
The current state-of-the-art of synthetic models such as PHOENIX (\citealt{allard2012models}) combined with the recent advances in high-resolution NIR spectroscopy offer a new opportunity for the determination of the stellar parameters and elemental abundances of M dwarfs.

Some of the previous spectroscopic works such as \cite{rojas2010metal}, \cite{rajpurohit2018exploring}, \cite{marfil2021carmenes} and \cite{cristofari2022estimating} have characterized M dwarfs through high-resolution spectroscopy using synthetic models. However, the currently popular synthetic models, specially for cool stars like M dwarfs, are incomplete and, up to a noticeable level inaccurate \citep{blanco2019modern}. A few examples of such incompleteness in the current synthetic spectra are the lack of identification of many observed atomic and molecular features due to incomplete line lists and inaccurate line-formation for weak lines in the synthetic spectra (\citealt{onehag2012m}). High-resolution spectroscopy on M dwarfs is the first step to improving the theoretical base of synthetic spectra via a detailed comparison between the data and the models. 

This paper presents the chemical spectroscopy of Barnard's star observed by the \spirou instrument. This includes a high-resolution spectral analysis and examination of over 18000 absorption lines and molecular bands. We describe the observations with SPIRou in Section \ref{sec:observation}.  The remaining bulk of the paper focuses on three subjects. First, in Section \ref{sec:synthetic}, we examine the caveats of one of the best synthetic spectra for M dwarfs by giving a detailed comparison of the discrepancies between the data and the models. Second, in Section \ref{sec:spectral}, we present a new method for determining the effective temperature and metallicity of M dwarfs that minimizes uncertainties inherent to the PHOENIX-ACES synthetic models. In Section \ref{sec:discussion}, we report the chemical abundance of 15 different elements for Barnard's star, and discuss potential reasons behind discrepancies in the reported abundances in the literature. This analysis is followed by concluding remarks in Section \ref{sec:summary}.

\begin{deluxetable}{ccc}
\tablecaption{Barnard's star (Gl 699) stellar properties}
\tablehead{
\colhead{Parameter} & \colhead{Value} & \colhead{Ref.}
}
\startdata
\multicolumn{3}{c}{\textit{Designations}}\\
TIC & 325554331 & 1                      \\
2MASS & J17574849+0441405 & 2            \\
UCAC4 & 474-068224 & 3                   \\
\textit{Gaia} DR3 & 4472832130942575872 & 4\\
\hline
\multicolumn{3}{c}{\textit{Astrometry}}                        \\
RA (J2016.0) & 17:57:48.50 & 4                                 \\
DEC (J2016.0) & +04:41:36.11 & 4                               \\
$\mu_{\alpha} \cos \delta$ (mas/yr) & $-801.551$ $\pm$ 0.032 & 4 \\
$\mu_{\delta}$ (mas/yr) & 10362.394 $\pm$ 0.036 & 4            \\
$\pi$ (mas) & 546.9759 $\pm$ 0.0401 & 4                        \\
Distance (pc) & 1.8282 $\pm$ 0.0001 & 4                      \\
\hline
\multicolumn{3}{c}{\textit{Stellar parameters}}\\
\teff (K) & 3231 $\pm$ 21 & 5          \\
$\left[ {\rm M/H} \right]$ & $-0.48$ $\pm$ 0.04 & 5\\
$\left[ {\rm Fe/H} \right]$ & $-0.39$ $\pm$ 0.03 & 5\\
SpT & M4.2 & 6                        \\
$M_{\star}$ (M$_{\odot}$) & 0.159 $\pm$ 0.016 & 6 \\
$R_{\star}$ (R$_{\odot}$) & 0.1869 $\pm$ 0.0012& 6\\
\logg (dex) & 5.08 $\pm$ 0.15 & 7\\
$L_{\star}$ (L$_{\odot}$) & 0.00342 $\pm$ 0.00003 & 6\\
\hline
\multicolumn{3}{c}{\textit{Photometry}}\\
$V$ & 9.540  $\pm$ 0.031 & 8        \\
$R$ & 8.315  $\pm$ 0.012 & 8        \\
$I$ & 6.730  $\pm$ 0.020 & 8        \\
$g$ & 10.428 $\pm$ 0.020 & 8       \\
$r$ & 8.913  $\pm$ 0.011 & 8       \\
$i$ & 7.508  $\pm$ 0.013 & 8       \\
$J$ & 5.244  $\pm$ 0.020 & 2        \\
$H$ & 4.834  $\pm$ 0.034 & 2        \\
$K_{\rm s}$ & 4.524 $\pm$ 0.020 & 2 \\
\multicolumn{3}{c}{\textit{Rotation}}\\
Rotation Period (days) & 145 $\pm$ 15 & 9 \\
$v$\,sin\,$i$ (kms$^{-1}$) & $<$\,2 & 10 \\
\enddata
\tablerefs{1.\ TIC \citep{stassun2019revised} 2.\ 2MASS \citep{skrutskie2006two} 3.\ UCAC4 \citep{zacharias2013fourth} 4.\ \textit{Gaia}  EDR3 \citep{vallenari2021gaia} 5.\ This work 6.\ \cite{mann2013spectro} 7.\ \cite{maldonado2020hades} 8.\ Synthetic photometry \citep{mann2015constrain} 9.\ \cite{toledo2019stellar} 10. \cite{reiners2018carmenes}}
\label{table:stellarparams}
\end{deluxetable}

\section{Observations}  \label{sec:observation}

\begin{figure*}
    \centering
    \includegraphics[width=\textwidth]{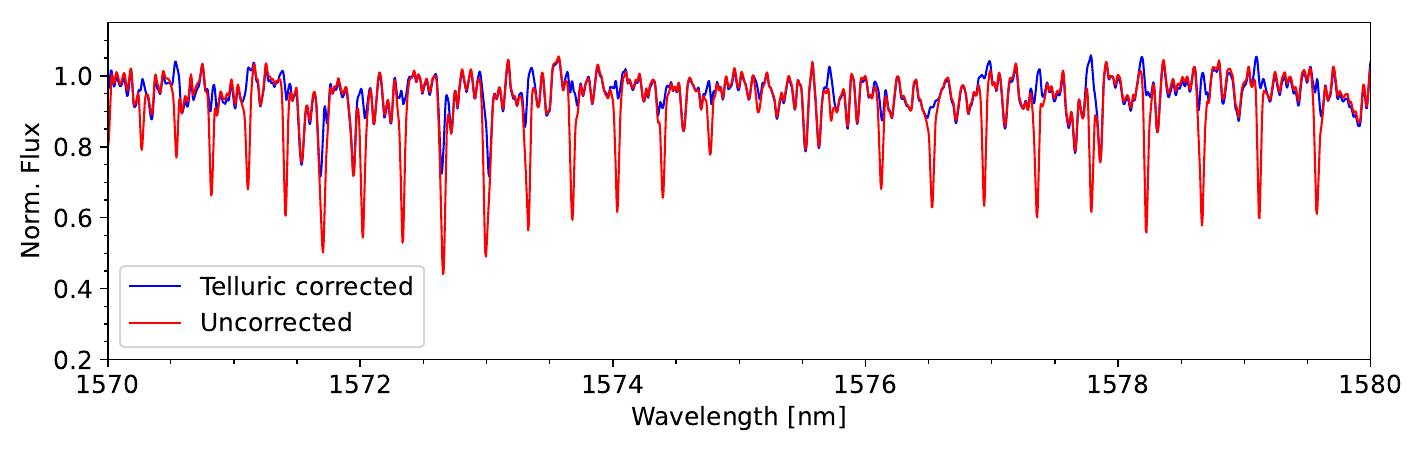}
    \caption{Spectral comparison showcasing the impact of telluric correction for a single visit. The red spectrum illustrates the uncorrected spectrum, while the blue one represents the spectrum after telluric correction using APERO (\citealt{cook2022apero}).}
    \label{fig:telluric}
\end{figure*}

\spirou (SpectroPolarimètre InfraRouge, \citealt{donati2020spirou}) is a near-infrared spectropolarimeter on the Canada-France-Hawaii Telescope (CFHT) that provides simultaneous high-resolution (R$\sim$70000) NIR spectra in the $YJHK$ bands between 0.97 and 2.49\,$\mu m$. This wavelength interval includes some of the common spectral metallicity indicators in M dwarfs (e.g., Ca I triplet and Na I doublet; \citealt{rojas2010metal}; \citealt{veyette2016physical}) as well as hundreds of known absorption lines that can be used for better determination of metallicity, \logg and \teffns. \spirou was specifically designed for the detection of exoplanets via precision velocimetry (1--2\,m/s) with unique polarimetric capability that enables measurements of the surface magnetic field of stars. Barnard's star (Gl 699), one of the closest and brightest M dwarf ($d=1.8$\,pc), is one of \spirouns's radial velocity standards regularly observed as part of the \spirou Legacy Survey\footnote{The SLS is one of CFHT's large programs and it focuses on exoplanet detection and characterization and magnetic fields of young M stars.} (SLS, \citealt{donati2020spirou}). The stellar properties of Barnard's star are given in Table \ref{table:stellarparams}.
The spectrum presented here results from the median co-addition of 846  visits secured between 2018 and 2023, enabling exceptionally high SNR ($\sim$1000).  Each visit consisted of a polarimetric sequence comprising four consecutive 60\,s exposures each observed in a different polarization state. 
The observations were reduced using APERO (A PipelinE to Reduce Observations; \citealt{cook2022apero}) that provides full calibration, extraction, and telluric correction, enabling corrections at the m/s precision level, with the maximum telluric residual of $< \,$1\% of the continuum level in the combined spectrum (see Figure \ref{fig:telluric}).

Of prime importance in our analysis is the spectral fidelity on a spectral scale of a few resolution elements. A spurious telluric residual signal could affect a line measurement for example. The spectrum analyzed here is the median combination of observations obtained at barycentric velocities spanning the full yearly excursions for that star ($\pm26.5$\,km/s) and thus any telluric residual would be practically eliminated from the final spectrum.

\begin{figure*}
    \centering
    \label{fig:Figure1}
    \includegraphics[width=\textwidth]{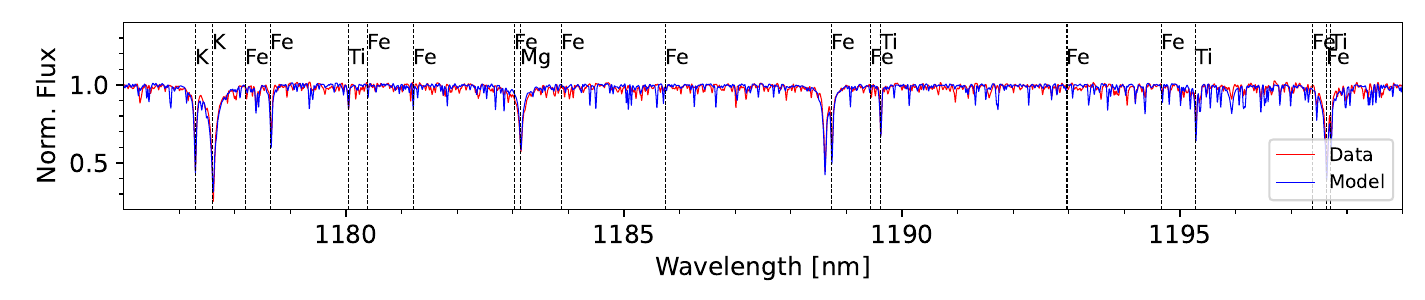}
    \includegraphics[width=\textwidth]{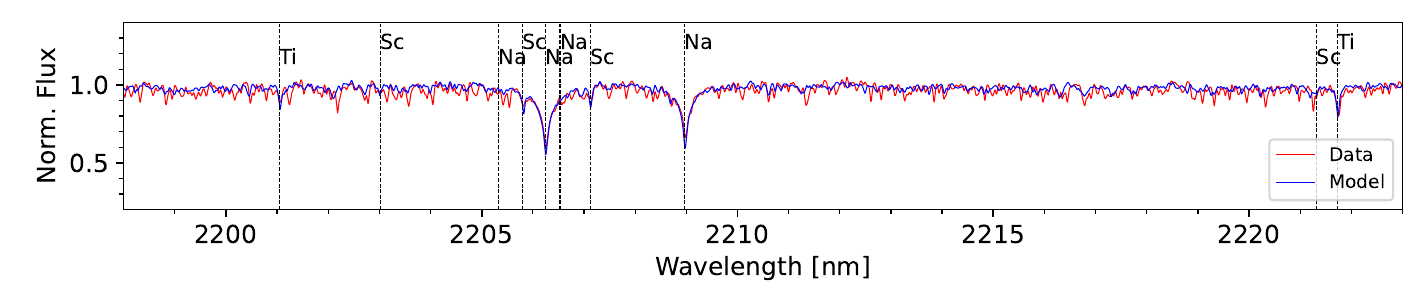}
    \includegraphics[width=\textwidth]{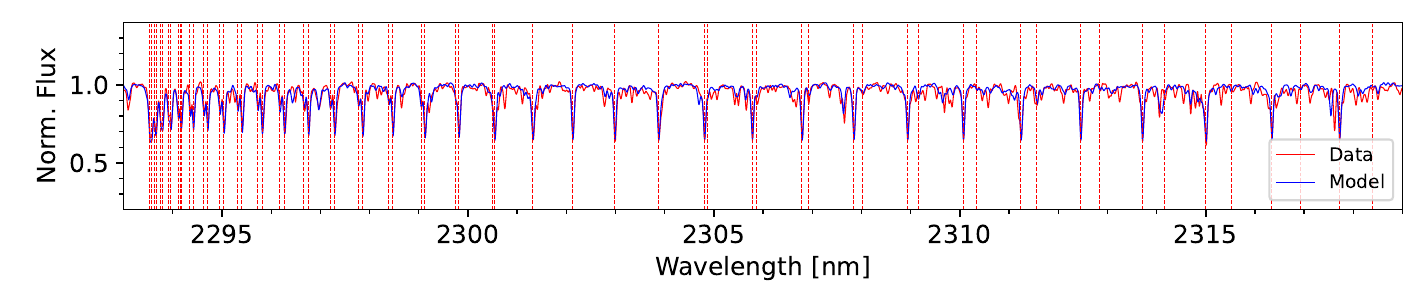}
    \caption{Comparison between the observed data (red) and the model (blue) with \teffns, \logg, and metallicity of 3200\,K, 5.0 dex and $-$0.5 dex, respectively. The \textit{Top panel} shows some alpha elements (e.g., Ti I and Mg I) as well as several isolated and blended Fe I lines.  The \textit{middle panel} shows the Na I doublet frequently used in several studies as metallicity indicators of M dwarfs. The \textit{bottom panel} shows numerous absorption features of the CO band head starting around 2.29\,$\mu$m (red dashed lines).
    }
\end{figure*}

\section{Synthetic Spectra} \label{sec:synthetic}

The chemical spectroscopic analysis of Barnard's star's spectrum was carried out using the PHOENIX-ACES models\footnote{PHOENIX version of 16.01.00B, released on 2012-02-11} (\citealt{allard2012models}; \citealt{husser2013new}). These synthetic spectra are generated using the pre-computed model grids from the PHOENIX radiative transfer code (\citealt{allard1996model}; \citealt{hauschildt1997parallel}; \citealt{allard2003new}). PHOENIX models are based on certain radiative transfer assumptions such as the convection process via the mixing length theory, hydrostatic equilibrium, and Local Thermodynamic Equilibrium (LTE) systems (\citealt{allard1996model}). By comparing synthetic spectra from the PHOENIX grid, we find that these models resemble our observed spectrum reasonably well\footnote{For further details, see Figures \ref{fig:full_comp1} to \ref{fig:full_comp5} in the appendix for the full spectral comparison, and Table \ref{table:full_data} containing the normalized fluxes of the observed data and a synthetic model analogous to Barnard's star.} (see Figure \ref{fig:Figure1}) except for the discrepancies described in Section \ref{subsec:Comparison}. This includes an excellent reproduction of both isolated absorption lines and molecular bands such as OH, CO, and CN. 

The spectral sampling of PHOENIX-ACES synthetic spectra varies between 0.01-0.04 {\AA} from the $Y$ to the $K$ band. We convolve PHOENIX-ACES models to the same spectral resolution of \spirou by creating an instrumental Line Spread Function (LSF). This LSF is then applied across a range of velocity bins, with the LSF recalculated for each bin to ensure accurate representation of variations across the wavelength grid. This process effectively matches the spectral resolution of the models to that of SPIRou.

Next, we utilized the iSpec tool (\citealt{blanco2014determining}; \citealt{blanco2019modern}) to normalize the convolved synthetic spectra. This tool first detects spectral peaks using a maximum filter, then filters out strong spectral lines and outliers, and further smoothens the data with a median filter. After these filtering steps, a cubic spline model is fitted to the continuum. This method effectively divided out the overall spectral energy distribution (SED), allowing for the isolation of specific spectral features. By performing this process all at once (and not order-based), a robust and consistent fit across the entire spectrum was ensured. The observed data was also normalized using the same method, ensuring consistency in the treatment of both synthetic and observed spectra.

\subsection{Comparison Between the Data and the Model} \label{subsec:Comparison}

While there is an overall consistency between the model and the data (see Figure \ref{fig:Figure1}), not all features in the observed spectrum exist in the synthetic model with \teff of 3200\,K, \logg of 5.0 and [M/H] of $-$0.5 dex, which are close to the stellar parameters of Barnard's star generally adopted in the literature (e.g., \citealt{mann2015constrain}; \citealt{artigau2018optical}). Our analysis (see Table \ref{table:line_statistics}) has revealed that there are over 18600 features in Barnard's star spectrum but only 6849 of those are identified in the model, assuming the following two selection criteria: 1) only lines with a minimum line depth of 5\% from the continuum level are considered and 2) the central wavelength of a given line measured from both the observed spectrum and the synthetic model are within one resolution element (assuming the spectral resolution of 70000)

Inconsistencies between the observed spectrum and synthetic model are discussed  below.

\subsubsection{Continuum mismatch} \label{Cont}

\begin{figure*}
    \centering
    \includegraphics[width=\textwidth]{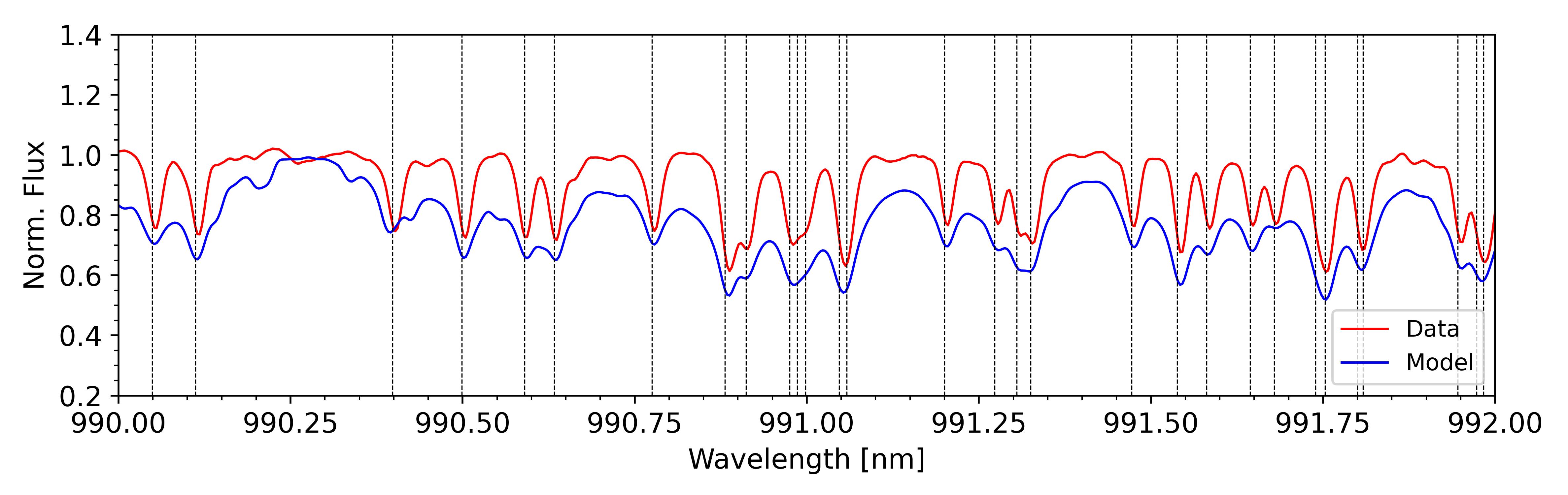}
    \caption{Example of a spectral domain with strong continuum discrepancy between the model (blue) and the data (red). The black dotted lines show the wavelengths of FeH lines. The same model as in Figure \ref{fig:Figure1} was used.}
    \label{fig:Figure2}
\end{figure*}

In about 5\% of the spectrum (mainly between 985-1068 nm), there is an apparent mismatch between the normalized continuum level of the data and the model. As shown in Figure \ref{fig:Figure2}, while the exact locations of FeH lines in the $Y$ band are projected correctly in the model, there is a noticeable difference between the continuum levels. A similar effect is also observed in \cite{lim2023atmospheric}, when they compared PHOENIX models with the JWST spectra. This discrepancy is not necessarily unique to the FeH bands, as there are some unaffected FeH lines in the $H$ band, but we empirically observed more severe mismatches around the molecular bands of the $Y$ band. This problem causes an overestimate for \teff and an underestimate for metallicity. Because of this issue, all spectral features between 985-1068 nm were excluded in the abundance analysis presented later.

\begin{figure*}
    \centering
    \includegraphics[width= \textwidth]{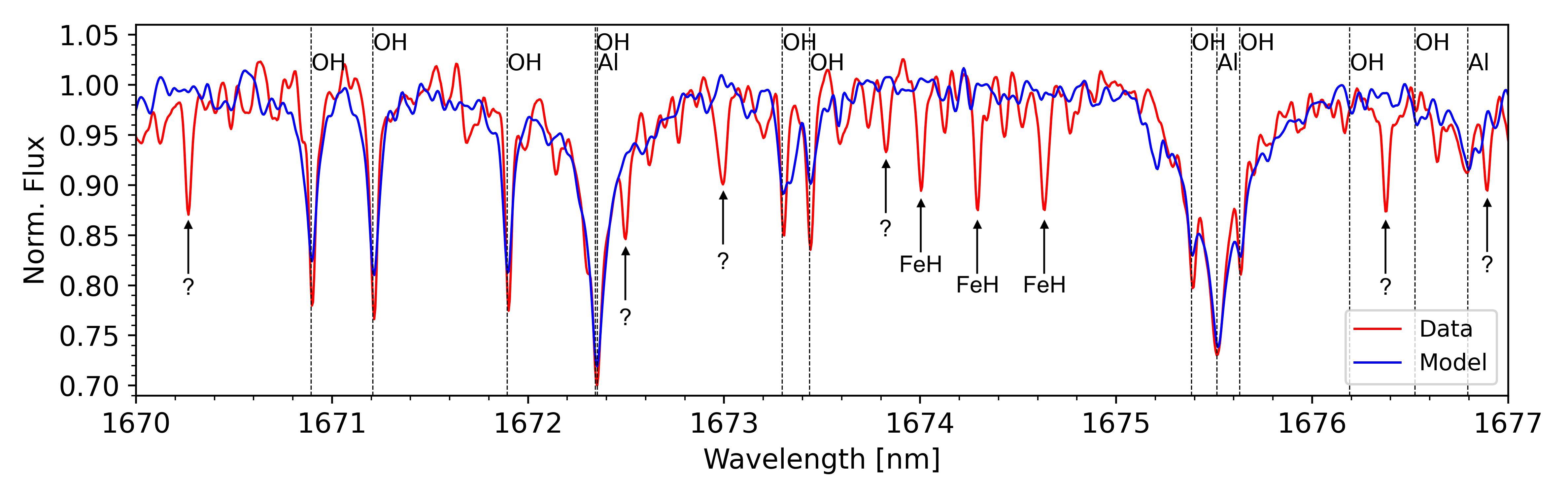}
    \caption{Example of a spectral region showing clear absorption features (red) not present in the model (blue).  Some of these lines, such as FeH lines, have been previously examined by \citealt{hargreaves2010high} and \citealt{souto2017chemical}.  The arrows with question marks show a few examples of unidentified lines out of thousands in the full wavelength range. The same model as in Figure \ref{fig:Figure1} was used.}
    \label{fig:Figure3}
\end{figure*}

\begin{figure*}
    \centering
    \includegraphics[width=1 \textwidth]{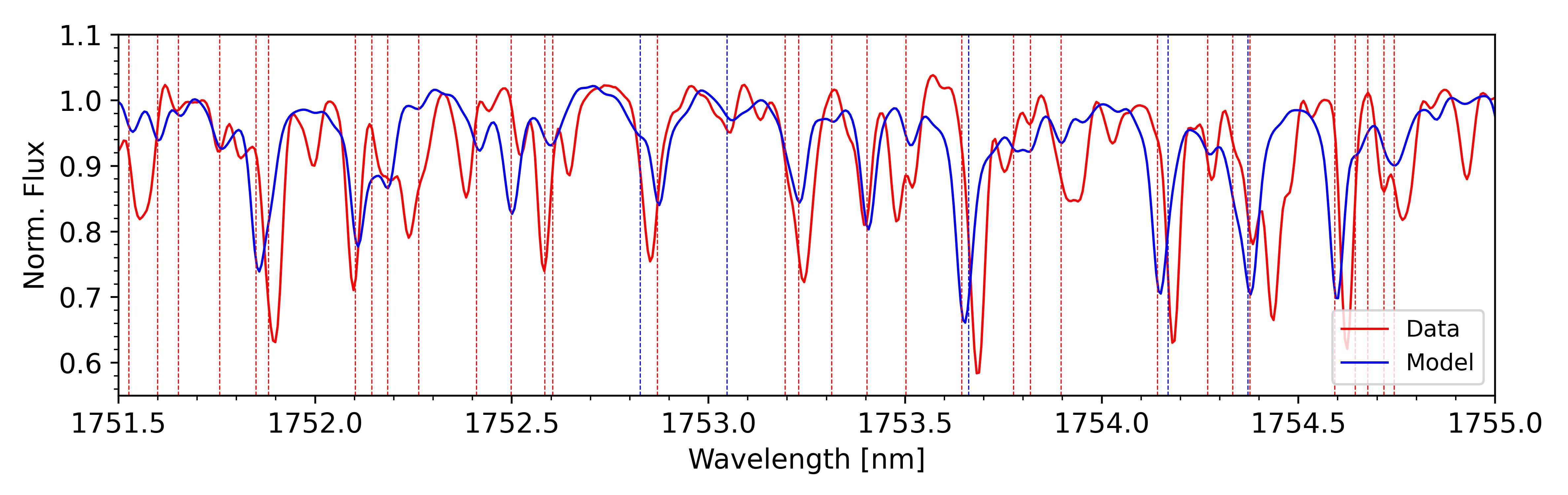}
    \caption{Spectral region featuring several water and OH lines.  There are multiple red and blue shifts between the data (red) and the model (blue).  The red and blue vertical dashed lines represent the location of H$_2$O and OH lines from the PHOENIX line list. The same model as in Figure \ref{fig:Figure1} was used.}
    \label{fig:Figure4}
\end{figure*}
\subsubsection{Unidentified lines in the models} \label{sec:unidentified}

Several thousand absorption features in the spectrum are absent from the default line list of PHOENIX (see Figure \ref{fig:Figure3}), hence these lines are not used in the model. While some of these unknown features were independently examined by different research groups (e.g., FeH lines by \citealt{hargreaves2010high} and \citealt{souto2017chemical}), or detected using the laboratory-based NIST line list database (\citealt{NISTAtomicSpectraDB}), there are still a significant number of unidentified spectral features in all the $YJHK$ bands.

\subsubsection{Line shifts} \label{sec:shifts}

As shown in Figure \ref{fig:Figure4}, some of the spectral features, usually associated with molecules, are shifted (typically less then 5 km/s). Similar discrepancy was also reported in other studies such as \cite{tannock20221}. A likely explanation for these shifts is an inherent wavelength uncertainty with the line list used in the synthetic spectra, however, this requires further investigations. These empirical shifts must be taken into account in the analysis for determining abundances and the effective temperature.  

\section{Spectral Analysis} \label{sec:spectral}

A $\chi^2$ minimization technique is the core of the spectral fitting procedure used in this paper. The stellar parameters (\teff and the overall metallicity) of the synthetic spectra  are varied and matched to the observed spectrum until convergence (minimal $\chi^2$) is reached. However, it is important to note that this $\chi^2$ minimization is not performed over the entire wavelength domain at once. Instead, the process is conducted iteratively over subsets of small spectral regions, the specifics of which will be defined later in the following sections.

In practice, multiple synthetic spectra from the PHOENIX-ACES grid were generated  with \teff between 2300\,K and  4300\,K (typical temperature range of M dwarfs) and with overall metallicity ranging from $-1.5$ to 0.5\,dex in 0.5\,dex increments. This metallicity range was chosen to cover the majority of the metallicities previously reported for Barnard's star; the range spans from -0.86 dex (\citealt{marfil2021carmenes}) to 0.61 dex (\citealt{passegger2021metallicities}).
Next, a fixed \logg of 5.0 dex was used for all our models, consistent with the mean and standard deviation of the reported \logg values for Barnard's star in SIMBAD (\citealt{wenger2000simbad}), 4.98\,$\pm$\,0.21\,dex.

Additionally we empirically confirmed that a variation in \logg within the range of 5.0\,$\pm$\,0.2 dex has a negligible effect on our analysis, and majority of the literature values for \logg are within this range. The relative insensitivity of the synthesized spectra to this parameter makes small offsets between the true and assumed values acceptable (well within our reported errors), thus the fixed value of \logg does not affect our results. Furthermore, by fixing \teff and \logg values determined independently, we can minimize additional uncertainties resulting from degeneracies in line shapes caused by various combinations of \teffns, \loggns, and metallicity.

\begin{figure}
    \centering
    \includegraphics[width=1\linewidth]{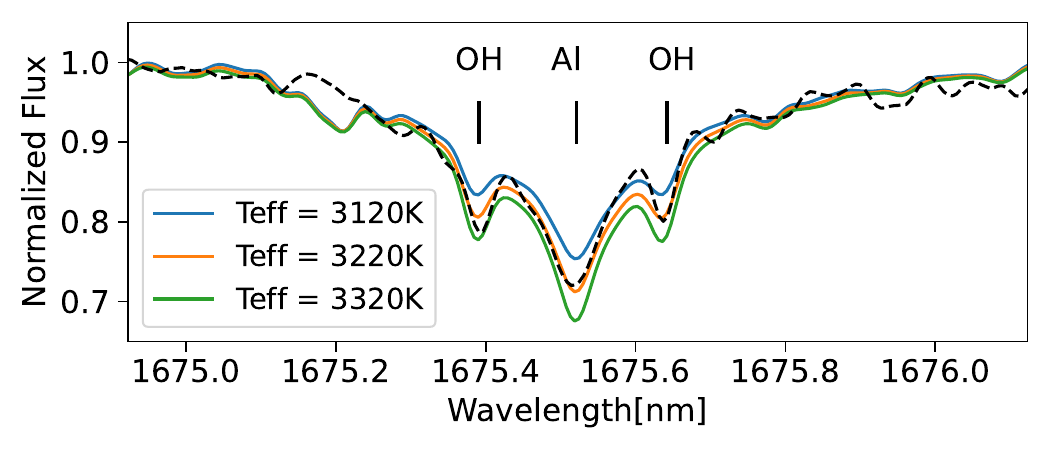}
    \includegraphics[width=1\linewidth]{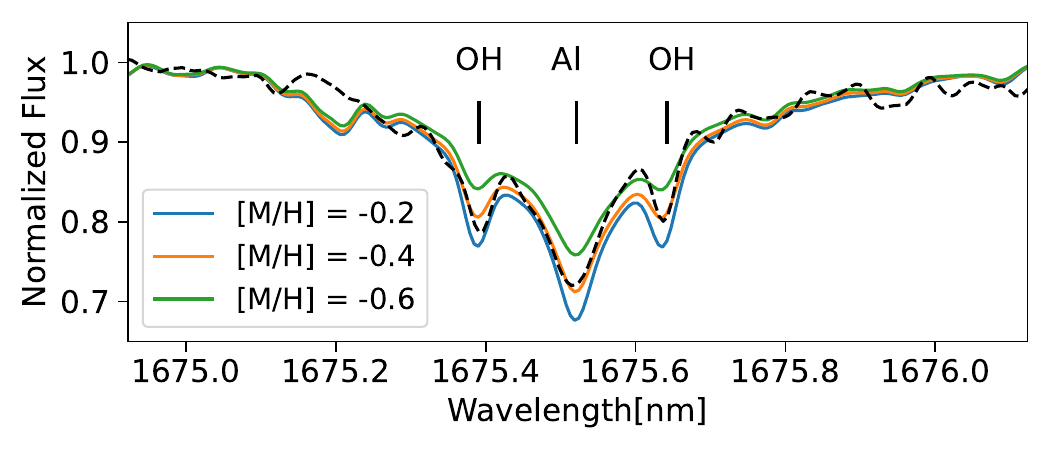}
    \caption{\spirou observation of the Al I line (1675.514\,nm) of Barnard's star (black dashed line). \textit{Top panel}: The solid lines represent the ACES models for a fixed metallicity of $-0.4$\,dex and \teff values of 3120\,K, 3220\,K, and 3320\,K.  \textit{Bottom panel}: Same as top panel, but the ACES models have a fixed \teff$ = 3220$\,K and metallicity values of $-0.6$\,dex, $-0.4$\,dex and $-0.2$\,dex. These plots illustrate the sensitivity of NIR high-resolution spectroscopy for constraining both the metallicity and effective temperature of M dwarfs.}
    \label{fig:Teff_met}
\end{figure}

The final $\chi^2$ fitting  was performed on a finer grid of models, with 20\,K and 0.1\,dex increments for the  \teff and metallicity, respectively, all  bi-linearly interpolated from the main pre-computed spectral grid (see Figure \ref{fig:Teff_met}).

\subsection{Line Selection}\label{subsec:Line}

\begin{deluxetable*}{llccc}
\tablecaption{Line selection statistics}
\tablehead{
& \colhead{Condition} & \colhead{Model} & \colhead{Data} & \colhead{Common lines}\\
\vspace{-0cm}& \colhead{} & \colhead{(\# of lines)} & \colhead{(\# of lines)} & \colhead{(\# of lines)}}

\startdata
 & Depth of +1\% cont. level            & 18926                    & 18617                   & 11155              \\
 & Depth of +5\% cont. level (criteria 1 \& 2, see Section \ref{subsec:Line})           & 13303                    & 12522                   & 6849               \\
 & Depth of +5\% cont. level (criteria 1 to 5, see Section \ref{subsec:Line}) - exist at \teff = 3200$^{\dag}$   & --                       & --                      & 1593               \\
 & Depth of +5\% cont. level (criteria 1 to 5, see Section \ref{subsec:Line}) - exist regardless of \teffns$^{*}$ & --                       & --                      & 636     \\   
& Depth of +5\% cont. level (all criteria in Section \ref{subsec:Line} and Section \ref{subsec:Abund})~$\hat{}\ $ & --                       & --                      & 210     \\   
\enddata
\tablecomments{
$^{\dag}$ The spectral features that exist in the data and a model with the effective temperature of 3200\,K, which is similar to Barnard's star's \teffns.\\
$^{*}$ The spectral features that exist in the data and models with the \teff range between 3000\,K and 4200\,K. This is the final list used for determining the \teffns.\\
$\hat{}\ $ This is the final list used for determining the chemical abundances.\\
}
\label{table:line_statistics}
\end{deluxetable*}

Line selection is a key component of any spectral analysis mainly for two reasons. First, as we have shown, since  synthetic models do not perfectly match the observed data, choosing spectral domains with good agreement between observations and models yields better results less susceptible to systematic effects. Moreover, different spectral features do not behave in the same way with the variation of physical parameters of the star, therefore, it is  important to choose spectral features that are highly sensitive to the changes in the spectral parameter that is being measured. For instance, OH lines are much less sensitive to effective temperature variation compared to H$_2$O molecular bands, meaning that for a fixed metallicity, the depth of the H$_2$O lines changes more significantly compared to OH lines for different \teff (\citealt{souto2017chemical}). Similarly, Fe I and Fe II lines have traditionally been used to constrain  \logg due to their different sensitivities to variations in \logg (\citealt{takeda2002spectroscopic}).

There are thousands of spectral features in the data and the synthetic spectra. Given the significant discrepancies between the model and the data (see Section \ref{subsec:Comparison}), blind spectral fitting to derive the stellar parameters and chemical abundances is not the best approach.
\begin{figure}[ht]
    \centering
    \includegraphics[width=1\linewidth]{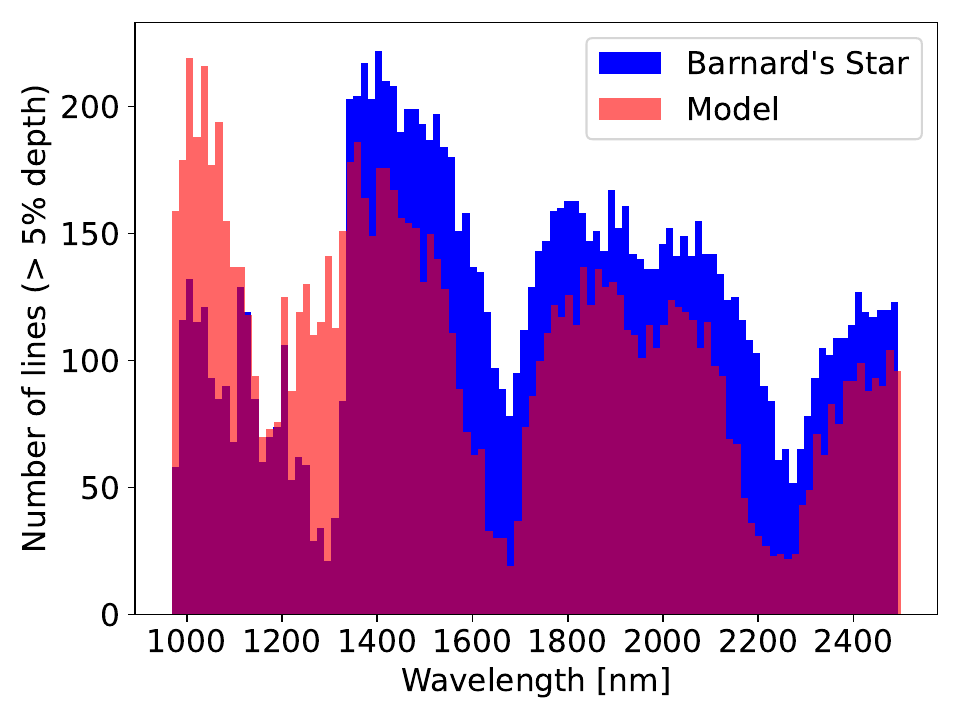}
    \caption{Comparison between the number of lines in the observed data (blue) and the synthetic model (red) with a similar \teffns, \logg and metallicity.}
    \label{fig:line_comp}
\end{figure}
As shown in Figure \ref{fig:line_comp}, and noted by \citet{artigau2018optical}, there are hundreds of spectral features missing in the $H$ and $K$ bands of a synthetic model with similar properties to that of Barnard's star. This is the reverse situation in the $Y$ and $J$ bands for which the model predicts numerous strong lines that are clearly not observed.  To avoid any bias toward the known and unknown caveats of the synthetic spectra, we developed a pipeline to find only those features that are common to both the data and the model. A spectral line is chosen only if it satisfies the following criteria:

\begin{enumerate}
    \item an observed line is present in the model
    \item the line depth is greater than 5\% from the continuum level\footnote{The telluric correction were done with the maximum residual level of well bellow 1\% in the combined spectrum, which is significantly smaller than our 5\% depth thresholds.},
    \item no continuum mismatch between model and observations (see Section \ref{Cont})
    \item the central wavelength from both the observed spectrum and the synthetic model do not differ more than one resolution element
    \item the lines that have a potential source of contamination or saturation (e.g.,nearby lines such as H$_2$O or OH lines), within half a resolution element, are flagged
\end{enumerate}

Using the above criteria, 1593 spectral features are identified from both the data and the synthetic model with \teff of 3200\,K, and \logg of 5.0 dex. However, for the purpose of \teff measurements, to avoid any line selection bias from the prior choice of 3200\,K for the model used for the line selection, we added another filter to pick only those lines that exist in the data and models regardless of the \teff of the star. More specifically, we selected the features that exist in models with \teff range between 3000\,K and 4000\,K and with line depth of 5\% or more. The final result is a total of 636 spectral features, that are in both the observed data and the synthetic model regardless of the stellar \teff (see Table \ref{table:line_statistics}).

\subsection{Group Determination of Effective Temperature} \label{subsec:Group}

\begin{figure}[ht]
    \centering
    \includegraphics[width=1\linewidth]{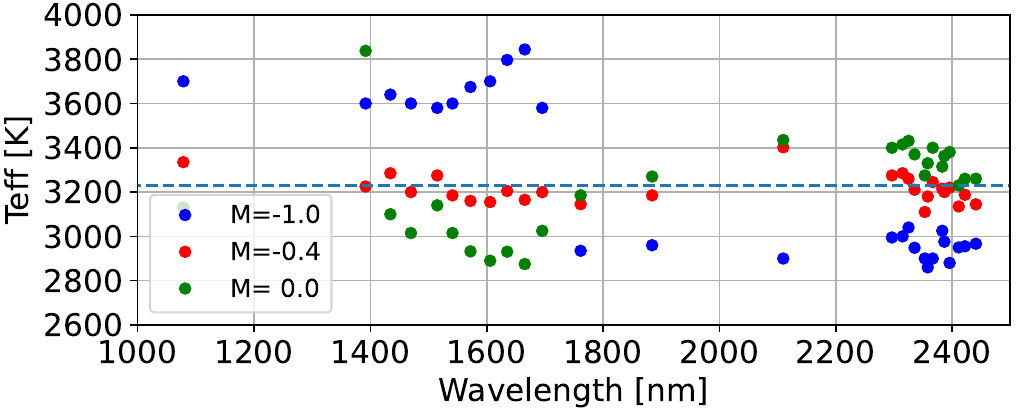}
    \caption{\teff vs wavelength for three different fixed overall metallicities. There is a strong correlation and anti-correlation between the estimated \teff for each scenario. The smaller the \teff variation is between different bands, the closer the model is to the data. The dashed line shows the average of \teff for the best fit.}
    \label{fig:Tscan}
\end{figure}

\begin{figure}[ht]
    \centering
    \includegraphics[width=1\linewidth]{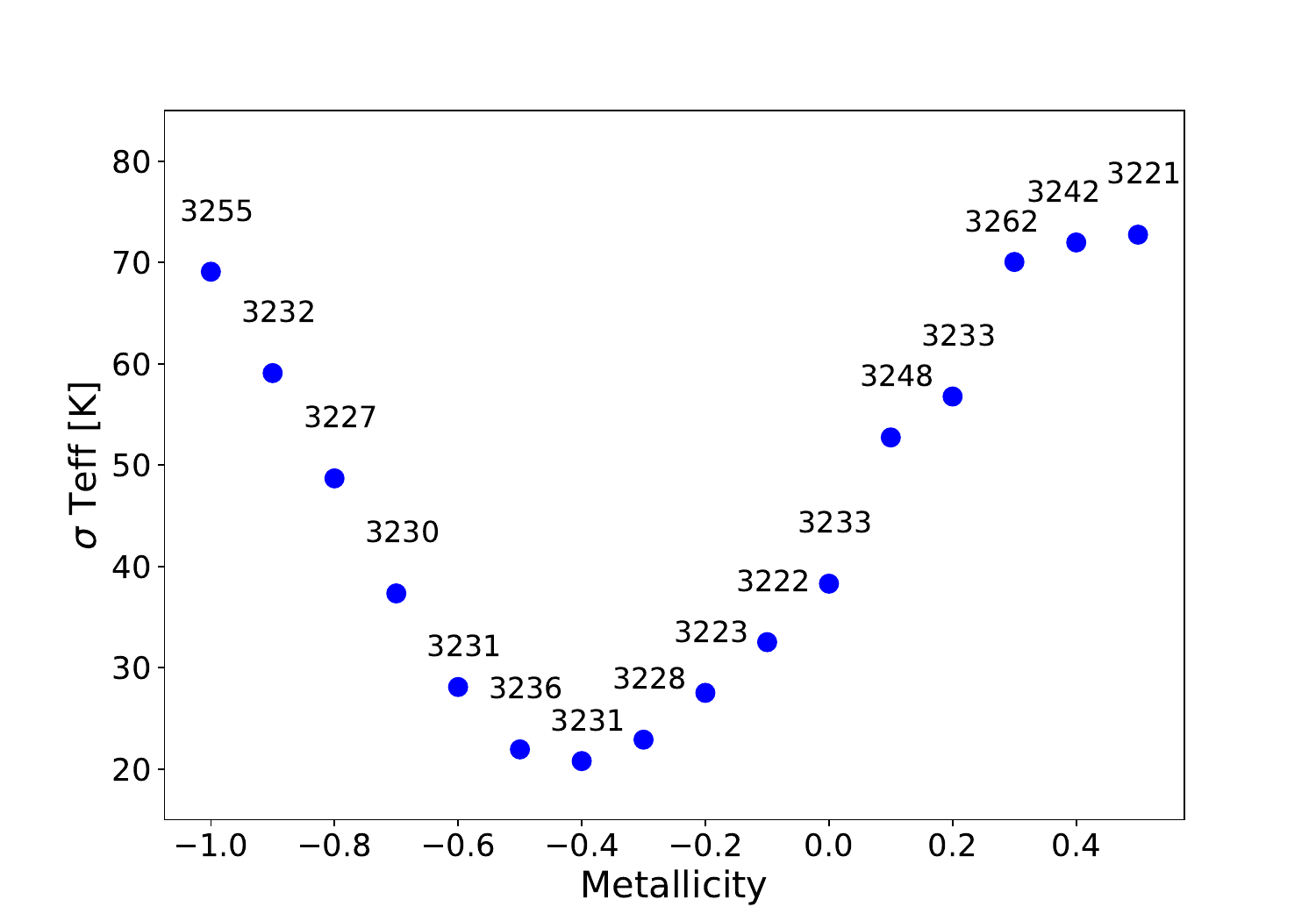}
    \caption{\teff dispersion for different fixed overall metallicities. The average \teff of each case is above each circle and its internal uncertainty is shown in the y-axis. The lowest variation corresponds to the metallicity of $-0.40\pm0.05$ and \teff of $3231\pm21$\,K. At this metallicity \teff values for different wavelengths have the lowest amount of variation between different spectral bands (see Figure \ref{fig:Tscan}). Note that the variation in \teff is calculated as the standard deviation divided by $\sqrt{N-1}$, where N represents the number of groups of individual lines used in this analysis.}
    \label{fig:stdev_metallicity}
\end{figure}

There is a significant variation of 3092\,K (\citealt{hojjatpanah2019catalog}) to 3463\,K (\citealt{fouque2018spirou}) for \teff of Barnard's star from previous studies. One of the plausible reasons behind such a variation in spectroscopic \teff is the choice of spectral features. We investigated this by a bulk line-by-line \teff determination on all 636 matched spectral features as a function of wavelength. In this method, we created multiple groups of spectral features that exist in both the data and the models, each containing several absorption lines or molecular bands (typically 20). It is important to note that these lines are not always adjacent to one another in the spectrum. Their specific separation can vary, depending on the location of the next matching line between the model and the data, and are not constrained to a fixed value.
We do not include the full spectral range from the first to the last line in a group; rather, our methodology involves a more targeted approach. Around each absorption line, we apply masking to isolate a subsection of the spectrum, specifically encompassing the closest local minima around each spectral line. This approach allows us to focus our analysis on the most relevant spectral features while avoiding potential noise or interference from less meaningful portions of the spectrum.
Then each group is analyzed independently through a $\chi^2$ fitting routine to infer both \teff and [M/H] for all groups. The heart of this method is determining \teff as a function of wavelength for different fixed metallicities. In each scenario, the metallicity is fixed to minimize the effect of unusually high or low abundance lines and only focus on the overall metallicity. In Figure~\ref{fig:Tscan}, three cases of \teff vs wavelength for the three fixed metallicities of $-1.0$, $-0.4$, and 0\,dex are shown. The \teff determined from $H$ and $K$ bands changes in opposite directions when the metallicity increases. We created a graph of \teff dispersion with different wavelength regions as a function of metallicity in Figure~\ref{fig:stdev_metallicity}.  The advantage of this method is that it yields several independent measurements that can be used to characterize the inherent uncertainties associated with the fitting procedure. As shown in Figure \ref{fig:stdev_metallicity}, the $\sigma_{T_{\rm eff}}$ of a given \teff shows a minimum with metallicity, allowing to constrain both \teff and [M/H]. This analysis applied to Barnard's star spectrum yields \teffns\,=\,3231\,$\pm$\,21 K and [M/H]\,=\,$-$0.40\,$\pm$\,0.05, in good agreement with the previous literature values (\citealt{mann2013spectro}; \citealt{gaidos2014trumpeting}; \citealt{gaidos2014m}; \citealt{maldonado2020hades}). Note that this is different from the overall metallicity listed in Table \ref{table:stellarparams}, that is determined via line-by-line fitting of different elements (will be discussed in Section \ref{subsec:Abund}).

Multiple points in this \teff line-by-line map have to be addressed. First, nearly all of the $Y$ and $J$ bands are excluded in this analysis partially due to the mentioned continuum mismatch of molecular bands (e.g., FeH bands) in these regions (see Section \ref{Cont}). The synthetic models systematically overestimate the depth of the majority of the FeH lines. Therefore, using the molecular bands in the $Y$ and $J$ bands causes a significant overestimation of \teffns. The reason is that for most of the $Y$ and $J$ bands, for a fixed metallicity, the hotter a star, the weaker the depth and equivalent width of molecular and atomic lines are. Therefore, the model has to increase the \teff to compensate for the continuum mismatch between the data and the model. Moving toward the $H$ and $K$ bands, the estimated \teff values are fairly consistent around 3200\,K.\\

\subsection{Abundance Determination} \label{subsec:Abund}

For this work, we generated a master line list by combining the most recent available line list used for Bt-Settl PHOENIX models (\citealt{allard2010model}) and the atomic database of the National Institute of Standards and Technology \citep{NISTAtomicSpectraDB}. These line lists contain a collection of atomic and molecular transition parameters, including the exact wavelength location of various atoms and molecules. While we used all 636 spectral features for the \teff determination, for the chemical abundance analysis, we took the conservative approach of selecting only atomic lines and OH lines. It is important to note that since the abundance of hydrogen is not modified when we change the overall metallicity of the models, the OH abundance from the best-fit model can represent the oxygen abundance. We excluded all non-OH molecular lines such as H$_2$O, FeH, TiO, and CO lines.
H$_2$O is very sensitive to \teff variations, unlike OH; therefore, minor inaccuracies in the fixed \teff can significantly misestimate any oxygen abundance inferred from H$_2$O lines. In section \ref{Cont}, we showed that our synthetic models suffer from continuum mismatch in most of the $Y$ band, which has a significant concentration of FeH lines. While there are some FeH lines in other bands, for consistency and due to the discrepancies with FeH lines in the $Y$ band, we decided to infer the iron abundance directly from the Fe I lines only. For other multi-metal molecules like TiO and CO, since we have control only over the overall metallicity in our synthetic models, properly disentangling the individual contributions of each molecule to the entire line is not feasible. Additionally, we removed ionized spectral features, as this work focuses solely on the chemical abundance of the neutral lines. Excluding all the mentioned spectral features, we selected 210 spectral features that are suitable for line-by-line chemical spectroscopy. Note that the wavelength spacing between two consecutive lines is often too small for the resolution of our spectrum (e.g., in some cases it was as small as 0.2 resolution element of our data). In these cases, we carefully flagged all known spectral features within half a resolution element of each line and labeled it as a ``Nearby Line" (see Appendix \ref{sec:appendix}).

\begin{figure}[ht]
    \centering
    \includegraphics[width=1\linewidth]{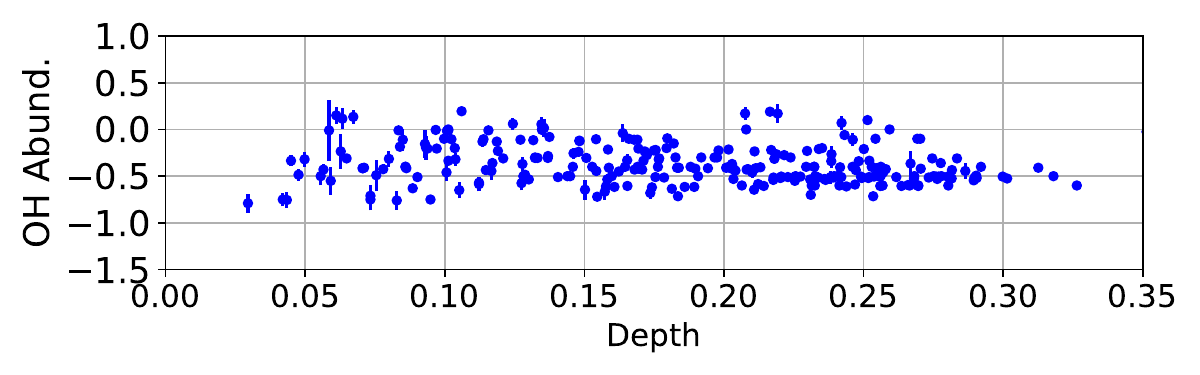}
    \caption{OH abundance as the function of line's depth. There is a correlation between each measurement's uncertainty (determined via MC sampling) and the depth of the line.}
    \label{fig:OH-depth}
\end{figure}

To minimize the human bias in the selection and analysis process, all line selection criteria described above are applied automatically. However, a few of the weaker lines still needed direct supervision for further confirmation to use them in the analysis. To make this quantifiable, we used all the available lines of hydroxyl molecules (OH) to determine the optimal threshold for the depth of the spectral lines that can be used in the automatic pipeline. By measuring the abundance of OH as the function of the depth of the lines (see Figure~\ref{fig:OH-depth}), we concluded  that all lines with a depth of less than 15\% from the continuum level require extra supervision as some of them are not reliable for precise chemical spectroscopy. This extra supervision includes confirmation of consistent continuum level of nearby lines between the data and the model.

\begin{figure}
    \centering
    \includegraphics[width=1\linewidth]{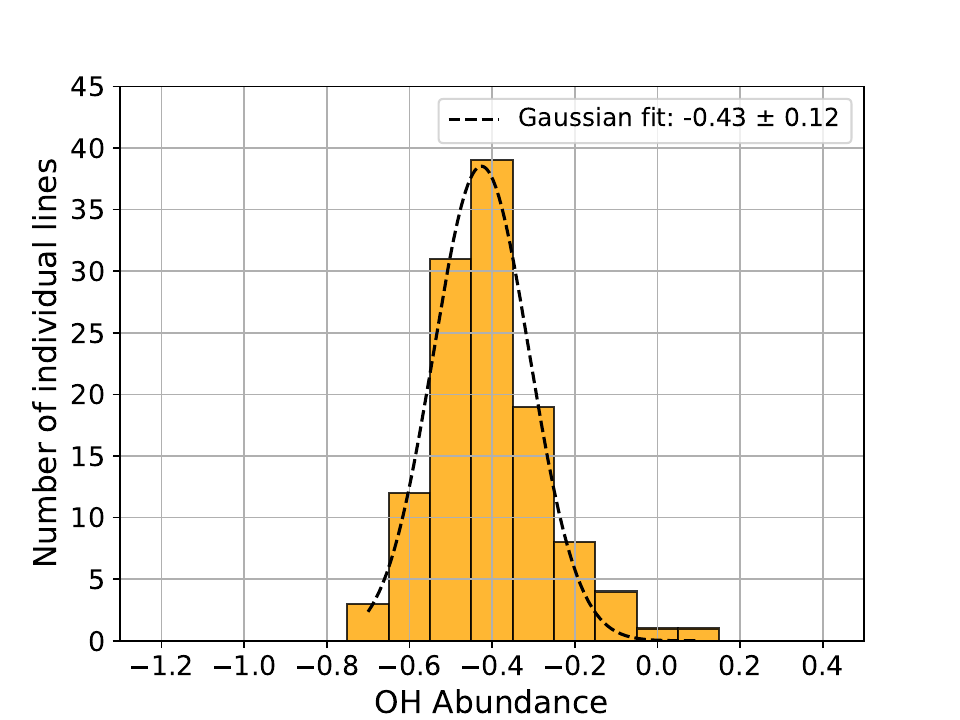}
    \caption{Distribution of OH abundance from 118 independent measurements of the OH lines that followed the five selection criteria of Section \ref{subsec:Line}.}
    \label{fig:oh-profile}
\end{figure}

\begin{figure}[ht]
    \centering
    \includegraphics[width=1\linewidth]{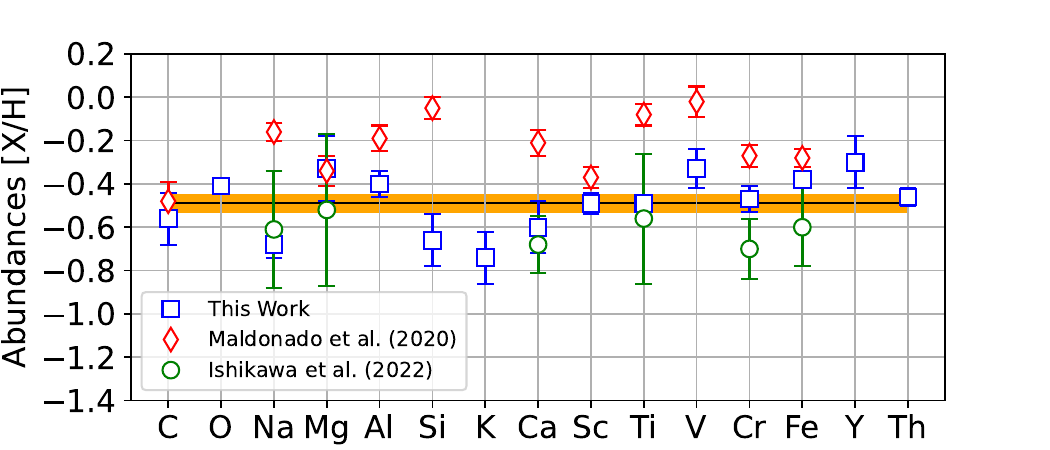}
    \caption{The chemical abundance of 15 different elements are determined via $\chi^2$ spectral fitting of our data using the PHOENIX-ACES synthetic spectra. The blue squares represent the solar-normalized [X/H] abundance of each element. The solid black line and its orange shade represent the overall metallicity and its corresponding uncertainty. The red diamonds and green circles represent the abundances from \cite{maldonado2020hades} and \cite{ishikawa2022elemental}, respectively.
    }
    \label{fig:abundances}
\end{figure}

The abundance of a given element was determined by averaging the abundances from all its lines, based on the best fit from synthetic models. The uncertainties are the standard errors derived from the dispersion in the data for elements with more than two lines, otherwise an uncertainty of 0.12  dex per line is adopted which is the value inferred empirically from the numerous OH lines (see Figure \ref{fig:oh-profile}). The solar normalized abundances of 15 different elements are reported in Table \ref{table:abundances} and Figure \ref{fig:abundances}. This work provides new abundance measurements for four elements: K, O, Y, and Th. 

In addition to the individual element abundances, two different integrated abundances are determined: the overall metallicity, [M/H], and the alpha abundance, [$\alpha$/H]. The overall metallicity ([M/H]) is defined as the average of the abundance of each element with the uncertainty expressed as the standard deviation of all values divided by $\sqrt{N-1}$, where N represents the number of different elements. This approach is chosen to avoid putting too much weight on the oxygen abundance characterized by a small uncertainty and to capture the observed dispersion from element to another. Note that for every spectral features, 50 Monte Carlo (MC) independent realizations of the observed spectrum are performed (see Appendix \ref{sec:appendix} for the MC errors) using the error for each pixel from the root-mean-square of the 846 spectra. This is done to better quantify the systematic uncertainty associated with our spectral fitting procedure. Since the spectrum of Barnard's star has a very high SNR, typical MC uncertainties are much smaller that the real dispersion inferred from several measurements of a given element.

\begin{deluxetable*}{ccccc}
\tablecaption{Stellar abundance of Barnard's star with respect to the Sun for various chemical species measured by \spirou}
\tablehead{
\colhead{[X/H]} & \colhead{This Work} & \colhead{\# lines}& \colhead{\citeauthor{maldonado2020hades}} & \colhead{\citeauthor{ishikawa2022elemental}} \\
\colhead{} & \colhead{} & \colhead{}& \colhead{\citeyearpar{maldonado2020hades}} & \colhead{\citeyearpar{ishikawa2022elemental}}}

\startdata
Fe I               & $-0.38 ~\pm ~$  0.03   &  29 & $-0.28 ~ \pm ~ 0.04$   & $-0.60 ~ \pm$ 0.18 \\
Mg I               & $-0.33 ~\pm ~$  0.15   &  3  & $-0.34 ~ \pm ~ 0.07$   & $-0.52 ~ \pm$ 0.35  \\
Ti I               & $-0.49 ~\pm ~$  0.03   &  18 & $-0.08 ~ \pm ~ 0.05$   & $-0.56 ~ \pm$ 0.30  \\
Cr I               & $-0.47 ~\pm ~$  0.06   &  8 & $-0.27 ~ \pm ~ 0.05$   & $-0.70 ~ \pm$ 0.14  \\
Na I               & $-0.68 ~\pm ~$  0.06   &  4  & $-0.16 ~ \pm ~ 0.04$   & $-0.61 ~ \pm $ 0.27 \\
Ca I               & $-0.60 ~\pm ~$  0.12   &  2  & $-0.21 ~ \pm ~ 0.06$   & $-0.68 ~ \pm $ 0.13 \\
Al I               & $-0.40 ~\pm ~$  0.06   &  4  & $-0.19 ~ \pm ~ 0.06$   & --                  \\
Si I               & $-0.66 ~\pm ~$  0.12   &  1  & $-0.05 ~ \pm ~ 0.05$   & --                  \\
C I                & $-0.56 ~\pm ~$  0.12   &  2  & $-0.48 ~ \pm ~ 0.09$   & --                  \\
Sc I               & $-0.49 ~\pm ~$  0.05   &  3  & $-0.37 ~ \pm ~ 0.05$   & --                  \\
V I                & $-0.33 ~\pm ~$  0.09   &  3  & $-0.02 ~ \pm ~ 0.07$   & --                  \\
K I                & $-0.74 ~\pm ~$  0.12   &  1  & --                     & --                  \\
O I$^*$            & $-0.41 ~\pm ~$  0.01   & 118 & --                     & --                  \\
Y I                & $-0.30 ~\pm ~$  0.12   &  1  & --                     & --                  \\
Th I               & $-0.46 ~\pm ~$  0.04   &  13 & --                     & --                  \\
$[$M/H]\,$\hat{}\ $   & $-0.49 ~\pm ~$  0.04   &  -- & --                     & --                  \\
$[\alpha$/M]$^\dag$   & $-0.01 ~\pm ~$  0.08   &  -- & --                     & --                  \\
\enddata
\tablecomments{$^*$The oxygen abundance is inferred from OH lines.\\
$\hat{}\ $Average abundance of all elements.\\
$^\dag$Average abundance of Mg I, Si I, Ti I, O I and Ca I alpha elements.}
\label{table:abundances}
\end{deluxetable*}

The alpha elements require special attention as they play a significant role in better understanding of not only the star itself but also the interior structure of their potential rocky exoplanet, in particular, the refractory elements such as Mg and Si,  that constitute the bulk material of a terrestrial exoplanet's core and mantle. In this work, we define the overall alpha abundance, or [$\alpha$/H], as the average abundance of all alpha elements detected in the spectrum, namely: Mg, O, Si, Ca and Ti. The Ti abundance is directly measured from the Ti I lines. We specifically avoided using the titanium oxide (TiO) spectral lines in the analysis to ensure that our results are not influenced or skewed by the contribution of oxygen in these molecules. The oxygen abundance is indirectly inferred from the OH absorption lines (see Figure \ref{fig:oh-profile}).
\begin{figure*}[ht]
    \centering
    \includegraphics[width=1\linewidth]{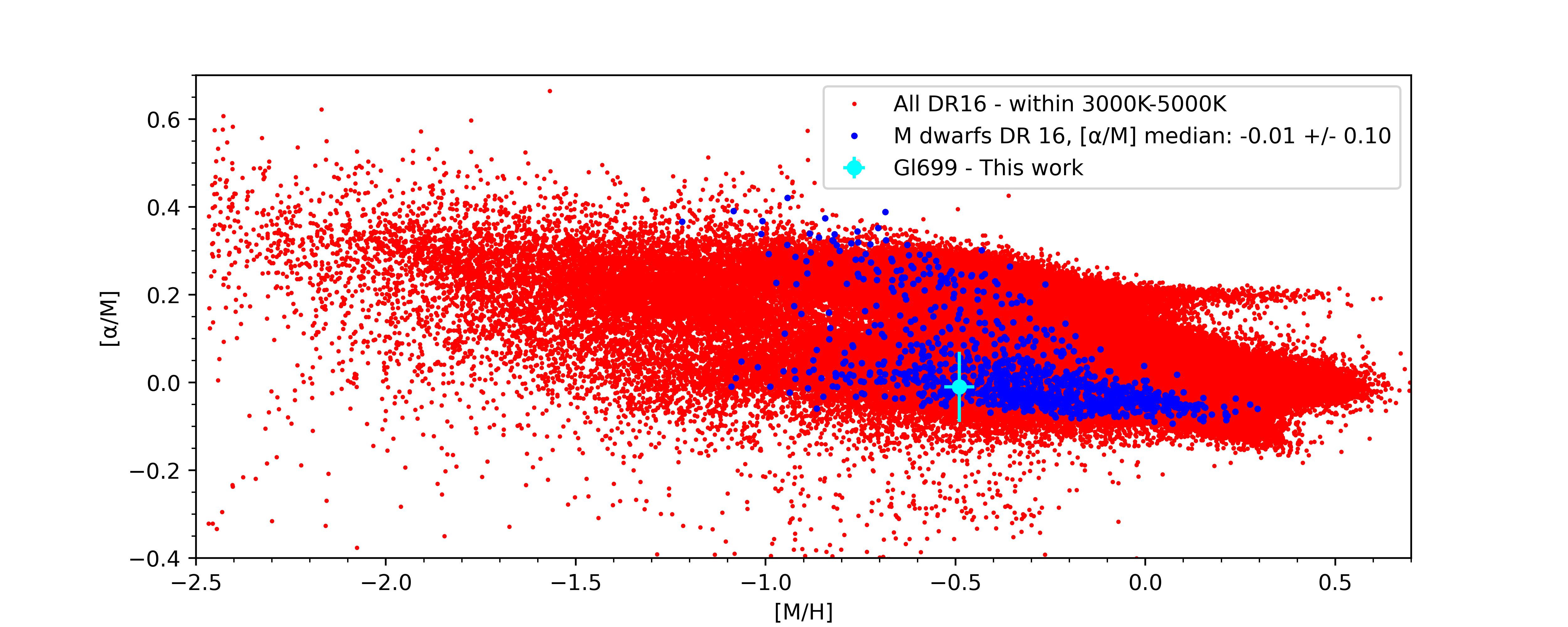}
    \caption{[$\alpha$/M] vs [M/H] from APOGEE DR16 for all different types of stars within \teff range of 3000 K to 5000 K (red circles) and M dwarfs (blue circles). The [$\alpha$/M] of Barnard's star from this work is highlighted with a cyan circle. While an anti-correlation is evident in the plot, not all M dwarfs have followed the same trend strictly.
    }
    \label{fig:mdwarfs_alpha}
\end{figure*}

The [$\alpha$/M] value determined in this work is $-$0.01\,$\pm$\,0.08 dex, which is lower than the typical 0.2\,$\pm$\,0.1 dex alpha abundance of metal-poor F and G stars in the thick disk (e.g. \citealt{bensby2014exploring}). However, recent observations from the APOGEE DR16 database (see Figure \ref{fig:mdwarfs_alpha}, \citealt{Majewski_2016}; \citealt{Ahumada_2020}) indicate that the trend of super solar alpha abundance in metal-poor stars may not be as pronounced for M dwarfs. Many M dwarfs in the APOGEE dataset, with metallicities similar to Barnard’s star, show [$\alpha$/M] values around 0 dex. This deviation is also observed by \citealt{ishikawa2022elemental}, where their Figure 13 distinctly illustrates that individual alpha elements vs metallicity in some M dwarfs are considerably lower than what is typically observed in thick disk FGK stars.

\section{Discussion} \label{sec:discussion}

Our analysis, based on the use of several groups of lines to determine the effective temperature, has yielded a \teffns\,=\,3231\,$\pm$\,21 K for Barnard's star, consistent with the value of 3238\,$\pm$\,11\,K inferred from the interferometric method (\citealt{mann2013spectro}), which is the most fundamental method for \teff determination but is more observationally expensive than spectroscopy. 

\begin{figure}[ht]
    \centering
    \includegraphics[width=1.\linewidth]{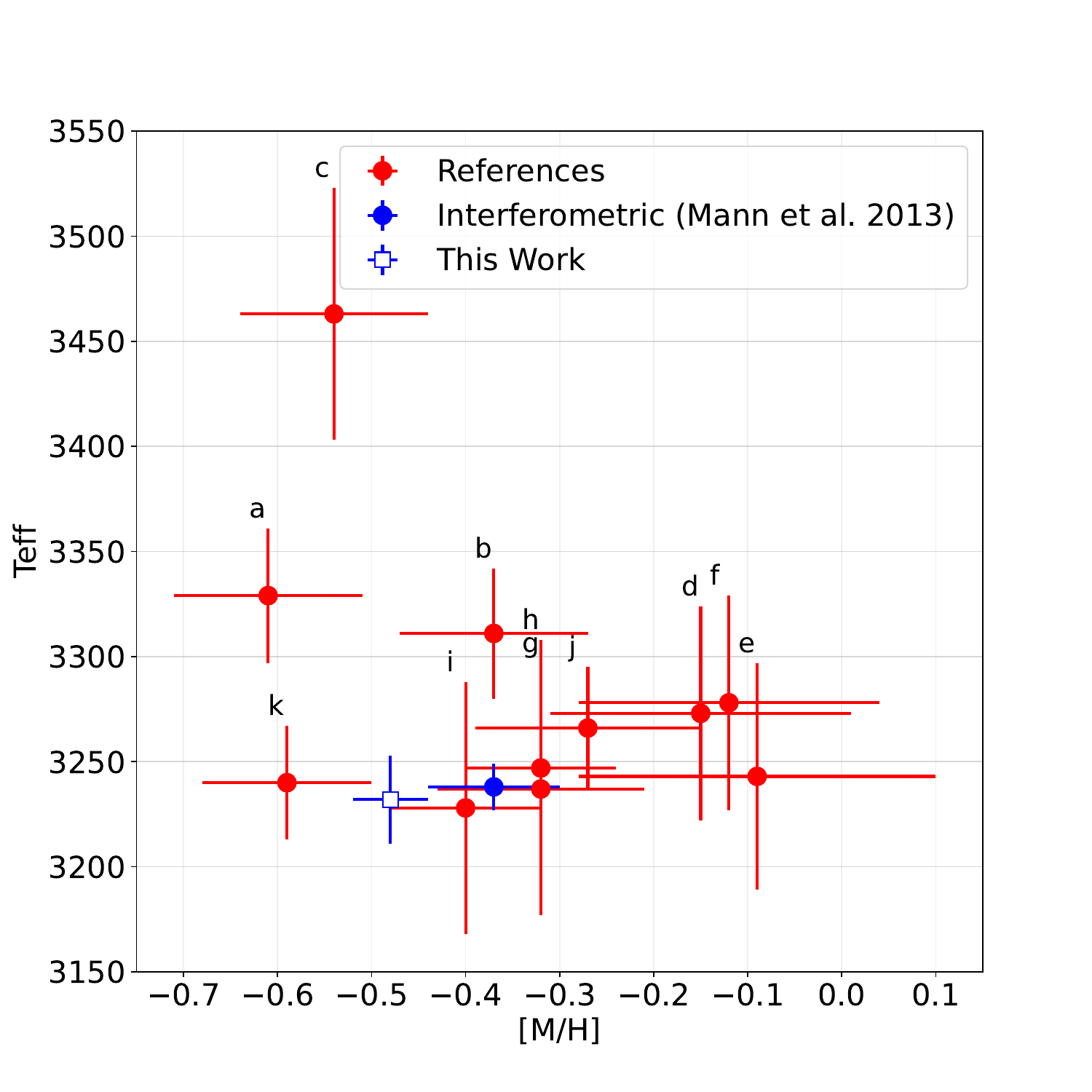}
    \caption{\teff and Metallicity of Barnard's star from different spectroscopic works. References: a.\,\cite{cristofari2022estimating} b.\,\cite{cristofari2022estimating2} c.\,\cite{fouque2018spirou} d.\,\cite{schweitzer2019carmenes} e.\,\cite{passegger2020carmenes} f.\,\cite{passegger2018carmenes} g.\,\cite{gaidos2014trumpeting} h.\,\cite{gaidos2014m} i.\,\cite{mann2015constrain} j.\,\cite{rojas2012metallicity} k.\,\cite{marfil2021carmenes}
    }
    \label{fig:Teff_references}
\end{figure}

As shown in Figure \ref{fig:Teff_references}, our \teff and [M/H] estimates show a fair agreement with most of previous works based on high-resolution spectroscopy but some significant variations are observed.  Indeed, it is interesting to note the \teff and [M/H] values inferred from this work differ significantly from \cite{cristofari2022estimating} even though both these analyses are based on the same \spirou dataset. 
In \cite{cristofari2022estimating}, four different \teff and metallicity values were presented, assuming various scenarios (e.g. fixed and variable \logg for PHOENIX and MARCS models), showing significant discrepancies due to the choice of synthetic models and line lists, leading to different results for the stellar parameters (the reported value in Figure \ref{fig:Teff_references} is for the case of fixed \logg with PHOENIX models that is similar to our analysis). Furthermore, \cite{fouque2018spirou} adopted a different approach by relying on equivalent-width measurements on ESPaDOnS spectra. These apparent discrepancies are likely related to the choice of different wavelength regimes and systematic effects resulting from a mismatch between observations and models, as shown in this work. As illustrated in Figure \ref{fig:Tscan}, not only does using the molecular features of $Y$ and $J$ bands (e.g., FeH) increase the probability of overestimating the \teffns, but different choices of metallicity and \teff can also interchangeably over- or under-estimate these parameters depending on the lines' wavelength regime. The effect of synthetic caveats varies depending on which part of the spectrum is used for the analysis, consequently impacting the determined fundamental parameters. Due to the group-fitting nature of our method, our work's estimated \teff is less sensitive to the choice of spectral feature; this methodology  provides a mean of calibrating other inherent uncertainties associated with a given choice of synthetic models.

\begin{figure*}[!htbp]
    \centering
    \includegraphics[width=0.45\linewidth]{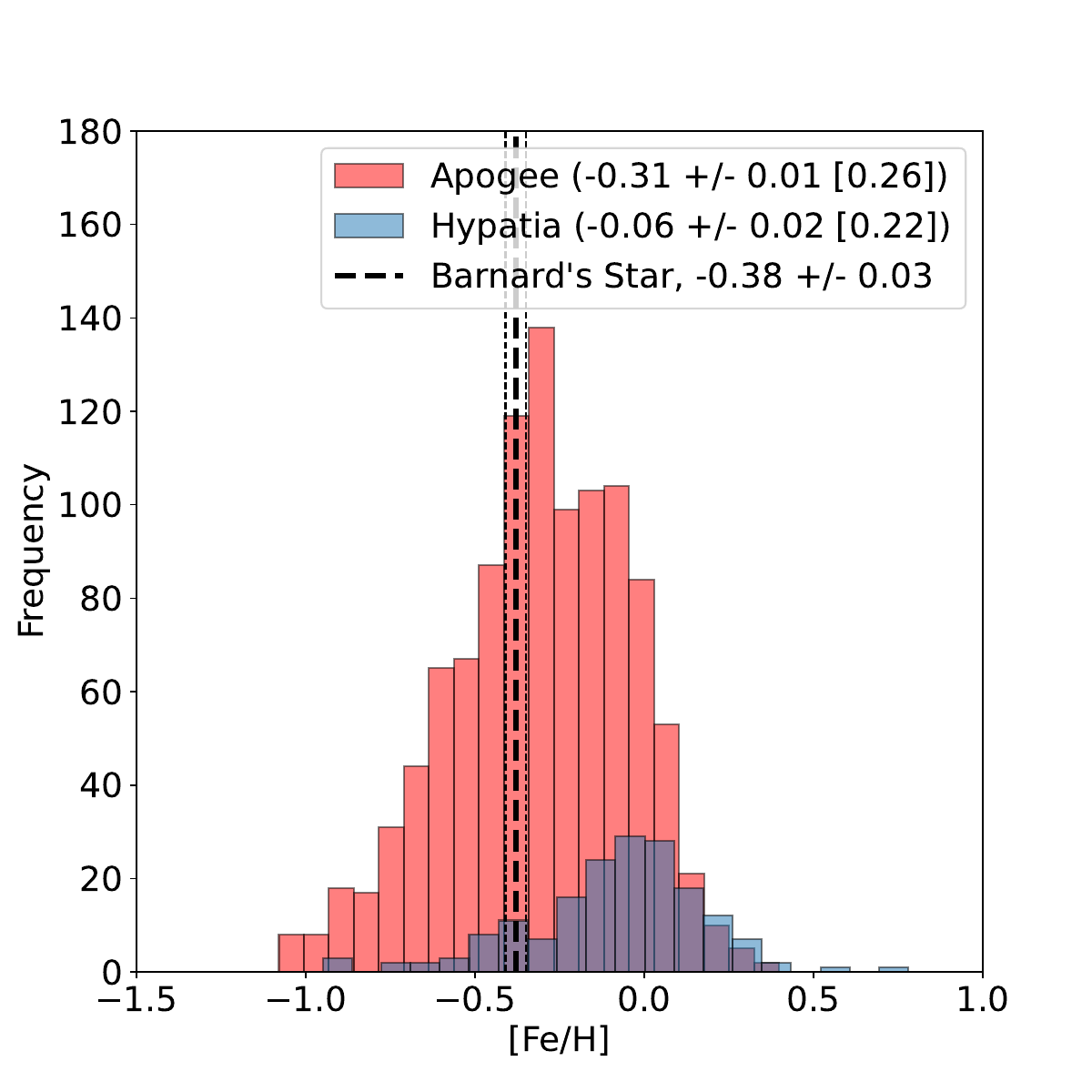}
    \includegraphics[width=0.45\linewidth]{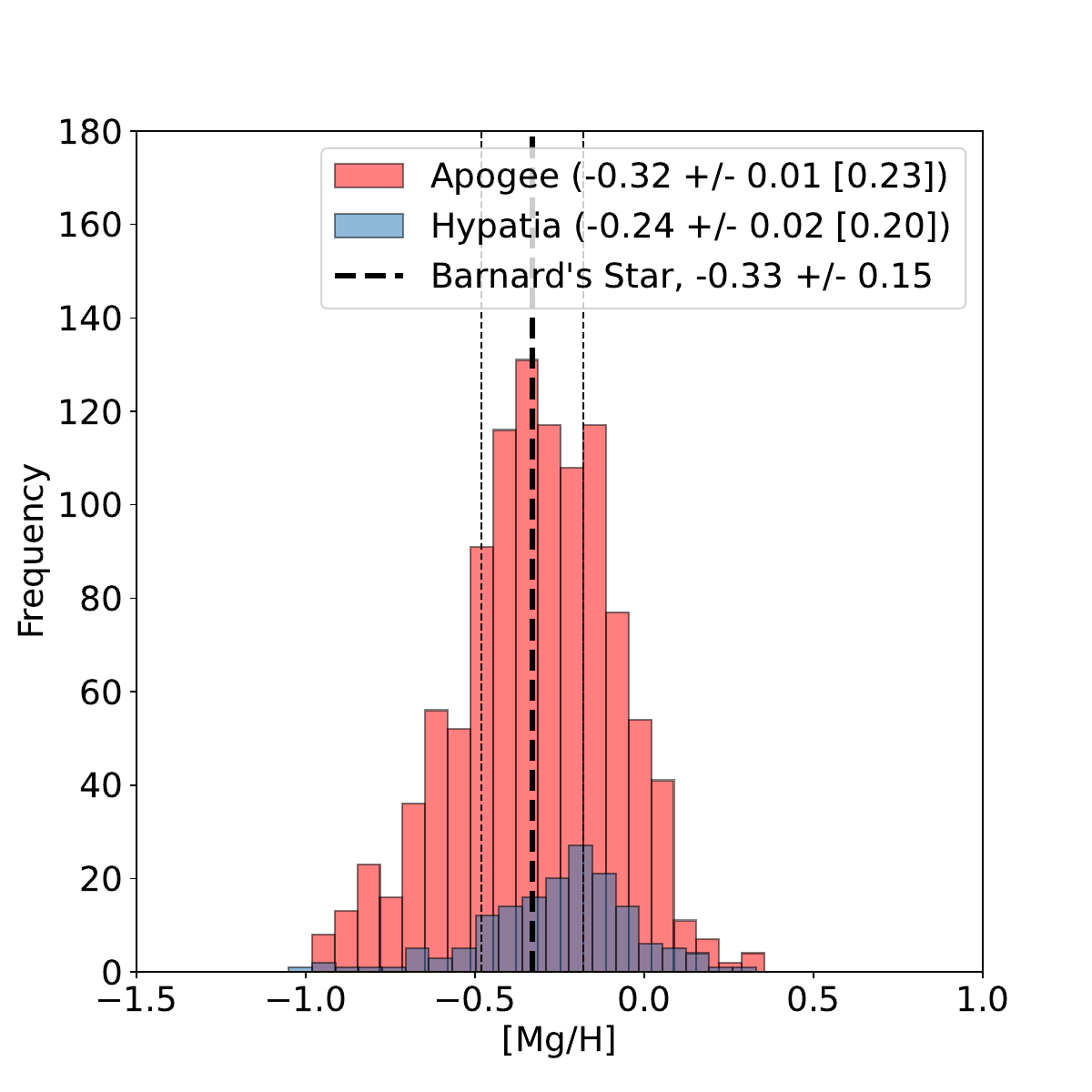}
    \caption{Comparison between the chemistry of M dwarfs from the APOGEE database (DR16) and Hypatia Catalog. The vertical dashed lines show Barnard's star's abundances from this work and their uncertainties. \textit{Left panel:} The [Fe/H] distribution of M dwarfs from both databases. Barnard's star's [Fe/H] from this work (the black vertical dashed line) is consistent with the APOGEE database but there is a noticeable tendency toward solar metallicity in the Hypatia Catalog. \textit{Right panel:} The [Mg/H] distribution of M dwarfs from both databases. The solar tendency in the Hypatia catalog is less severe for the Mg distribution, yet our result is still more consistent with the APOGEE abundances. The legend represents the mean $\pm$ relative uncertainty [standard deviation] for each database.
    }

    \label{fig:Met_comparisons}
\end{figure*}

By fixing the \teff and \logg in the models, we determined the chemical abundances of 15 different elements through $\chi^2$ fitting of 210 individual atomic lines or molecular bands. Figure \ref{fig:abundances} and Table \ref{table:abundances} provide a comparison of our abundance  measurements with that from the literature based on high-resolution spectroscopic observations. \cite{ishikawa2022elemental} used the high-resolution spectra from the IRD/Subaru Telescope, with their data covering a wide wavelength range from 0.97-1.75$\mu$m ($Y$, $J$, and $H$ bands), comparable to parts of the \spirou wavelength range. They determined the chemical abundances of several M dwarfs, including Barnard's star, by comparing the equivalent width of tens of spectral features with those of MARCS synthetic models. \cite{maldonado2020hades} developed a novel method to determine stellar abundances in M dwarfs using high-resolution optical spectra. They trained their model, which uses principal component analysis and sparse Bayesian methods, on M dwarfs orbiting FGK primaries. This model was then applied to a large sample of M dwarfs, including Barnard's star.

Our results are consistent with abundance measurements from  \cite{ishikawa2022elemental} with respect to the large uncertainties of their studies. Note that the adopted \teff and \logg in \cite{ishikawa2022elemental} are 3259\,$\pm$\,157\,K and 5.076\,$\pm$\,0.028\,dex, that are similar to our adopted value. This implies that a change in \teff alone would not have put these into better agreement. Our abundance measurements of Fe, Mg, C and Sc are within 2$\sigma$ of those of \cite{maldonado2020hades}, but others, most notably  Na and Si, are higher than our abundances. This comparison emphasizes the difficulty of inferring accurate abundance measurements of individual elements from high-resolution spectroscopy. It is difficult to identify the cause for such discrepancies but the use of the different synthetic models, in addition to the choice of lines, are likely the primary reasons.  

\subsection{Fe, C, Si, Mg and O} \label{subsec:Si}

\begin{deluxetable*}{ccccc}
\tablecaption{Abundance ratios for Barnard's star}
\tablehead{
\colhead{X/H} & \colhead{This Work} & \colhead{Sun$^\dag$} & APOGEE$^{*}$& Hypatia$^{*}$}

\startdata
C/O   & 0.39 $\pm$ 0.32              & 0.55     $\pm$ 0.17  & [0.25, 0.76] & --\\
Mg/Si & 2.63 $\pm$ 0.52              & 1.23     $\pm$ 0.12  & [0.87, 2.20] & [0.32, 1.83]\\
Fe/Mg & 0.71 $\pm$ 0.42              & 0.79     $\pm$ 0.14  & [0.38, 1.26] & [0.42, 2.22]\\
Fe/O  & 0.07 $\pm$ 0.08              & 0.07     $\pm$ 0.16  & [0.03, 0.09] & -- 
\enddata
\tablecomments{
$^{\dag}$ Photospheric abundance ratios from \citet{asplund2009chemical}.\\
$^{*}$95\% confidence interval of the M dwarf population ($\sim$1000) from APOGEE DR16 (\citealt{Majewski_2016}; \citealt{Ahumada_2020}) and ($\sim$140) Hypatia database (\citealt{Hinkel2014}).}
\label{table:ratios}
\end{deluxetable*}

Several studies have unveiled a correlation between the metallicty of extrasolar host stars and the occurence rate of gas giant exoplanets (\citealt{fischer2005planet}; \citealt{bond2006abundance}; \citealt{guillot2006correlation}). Specifically, these works have revealed that host stars are typically enriched in Fe, C, Si, Mg, and Al at various levels. These elements are the building blocks of planetary cores that lead to the formation of both rocky and giant planets. 
Since there is both empirical and theoretical evidence that stellar relative abundances of refractory elements are a good proxy for planets \citep{Dorn2017}, one can get some constraints on the chemical composition of planetary cores through stellar abundance measurements. The near-infrared spectrum provides several spectral lines for abundance measurements of refractory elements (i.e. Fe, Si, Mg) as well as for C and O. The C/O ratio in the atmosphere of gas giant exoplanets provides some constraint on the planet formation location within the circumstellar disk \citep{Oberg2011}.

Table \ref{table:ratios} gives the C/O, Mg/Si, Fe/Mg, and Fe/O ratios of Barnard's star compared to that of M dwarfs from the APOGEE database\footnote{The APOGEE database is an IR spectroscopic survey comprising hundreds of thousands of stars spanning the entire Galactic bulge, bar, disk, and halo. This survey covers the wavelength range of 1.5-1.7$\mu$m in the $H$ band, with a spectral resolution R=22500 (\citealt{gunn20062}). This database provides estimates for stellar parameters and abundances of various elements.} (DR16, \citealt{Majewski_2016}, \citealt{Ahumada_2020}) and the Hypatia Catalog (\citealt{Hinkel2014}). While Mg/Si is noticeably larger than the solar value, all ratios are within the 95\% confidence interval of the M dwarf population, with respect to their uncertainties.

It is also interesting to note that there are significant differences in the [Mg/H] and [Fe/H] distributions from APOGEE and Hypathia (see Figure \ref{fig:Met_comparisons}),  highlighting the fact that abundance measurements in the literature suffer from fairly large dispersion and, likely,  significant systematic uncertainties due to different methodology. For instance, the APOGEE database is derived from homogeneous measurements from the same instrument using a common analysis methodology while the Hypathia catalog is a collection of heterogeneous abundance measurements from various sources in the literature.  

\subsubsection{Contamination of Si I Lines} \label{subsec:contamination_si}

\begin{figure}[ht]
    \centering
    \includegraphics[width=1.0\linewidth]{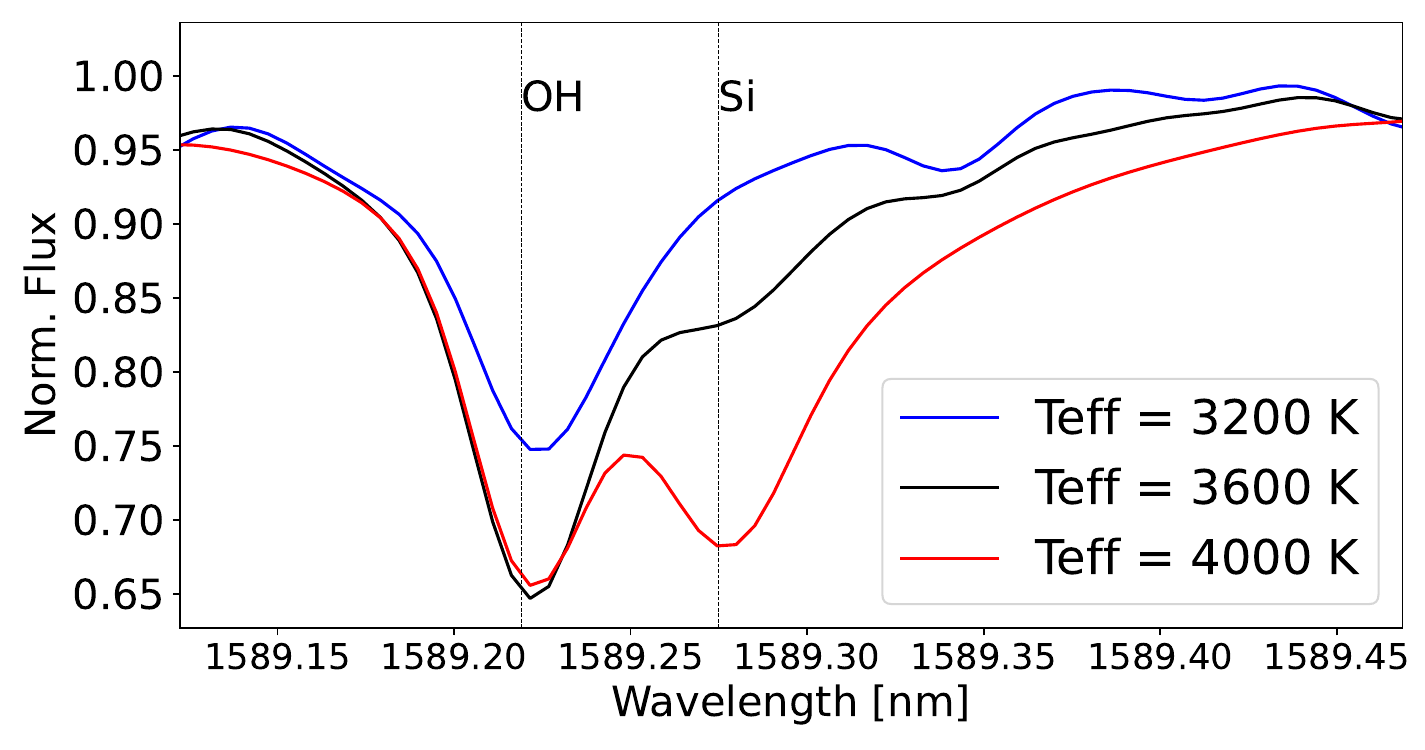}
    \caption{Comparison between three synthetic spectra from the PHOENIX-ACES model convolved to the resolution of SPIRou, with the same metallicity of $-$0.5 dex but different effective temperatures. This Si I line completely dissolves in the OH molecular line for temperatures lower than 3500\,K.}
    \label{fig:si_T_sens}
\end{figure}

By altering the \teff of a star, the depth and equivalent width of its spectral lines change nonlinearly at varying rates, depending on the nature of the spectral features. In most cases, molecular bands can either entirely suppress or contaminate the atomic wings or nearby weaker lines. This sensitivity is particularly significant for absorption lines that are inherently weak within the spectrum. Silicon lines serve as prime examples, illustrating the importance of considering temperature sensitivity in chemical spectroscopy. As an example, for a star with metallicity comparable to Barnard's star, a \teff of 3600\,K represents a threshold for precisely measuring the abundance of the commonly used Si I at 1589.27 nm (see Figure \ref{fig:si_T_sens}). Moreover, for spectra with a medium spectral resolution below 25000 (e.g., that of the APOGEE database), this specific Si line may fully or partially merge with its adjacent OH line, leading to a wrong estimation of the overall Si abundance.

\begin{figure*}[!tbp]
    \centering
    \includegraphics[width=1.03\linewidth]{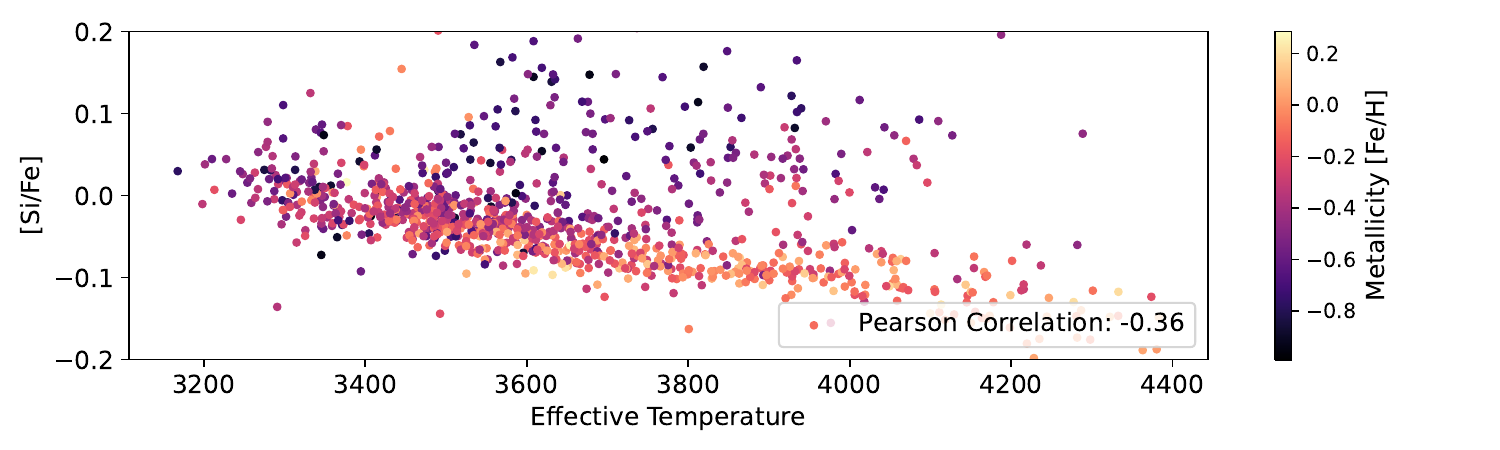}
    \caption{The silicon abundance of APOGEE M dwarfs with \logg range of 4.0 to 6.0 dex, and effective temperature range of 3100\,K to 4000\,K. There is an unexpected anti-correlation between the silicon abundance and \teffns.}
    \label{fig:si_Teff_apg}
\end{figure*}

To more effectively demonstrate the impact of temperature sensitivity on the chemical abundances of weak lines, we used the APOGEE database (\citealt{majewski2016apache}) to investigate the chemical behavior of M dwarfs at various temperatures and to identify any potential inconsistencies between the chemistry of our target and similar M dwarfs in the APOGEE database. This allows us to closely examine the Si abundance of M dwarfs with different effective temperatures.

Generally, no relationship is expected between an element's chemical abundance and a star's effective temperature. However, upon examining the chemical abundances of M dwarfs from the APOGEE DR16, a distinct anti-correlation is apparent in the [Si/Fe] distribution as a function of temperature (see Figure \ref{fig:si_Teff_apg}). This overestimation of Si abundance is more significant in colder and/or metal-poorer stars. This anti-correlation is likely due to the contamination of nearby strong lines on Si abundance measurements. For cooler stars, the OH lines fully dominate the adjacent weaker lines, while for hotter, metal-poor stars, the depth of the Si line becomes extremely weak, leading to a similar effect of merging with nearby lines. In either case, an overestimation of abundance is a probable outcome. We examined the existence of such an anti-correlation for other elements and discovered none or only minor comparable trends for the remaining elements. The conclusion from this exercise is that the Si abundance reported in APOGEE should be taken with caution specially for late M dwarfs.  

\section{Summary}\label{sec:summary}

Systematic caveats in synthetic models can significantly impact the chemical spectroscopy of stars. This work has highlighted some of the critical caveats that can lead to mis-estimation of fundamental stellar parameters, such as \teffns, and different chemical abundances for Barnard's star.

We examined the high-resolution spectrum of Barnard's star from the \spirou instrument via a $\chi^2$ analysis of multiple absorption lines using the PHOENIX-ACES synthetic spectra. We determined the effective temperature, overall metallicity and chemical abundances of 15 different elements. For the effective temperature, we employed a novel method based on the simultaneous group fitting of numerous spectral features, ascertaining the \teff by examining the sensitivity of these spectral features as a function of wavelength. 

To minimize the effects of uncertainties associated with synthetic models, we developed a pipeline that identifies common spectral features between observed spectra and synthetic models that are at least 5\% deep from the continuum level and are not saturated. Using the cleaned set of spectral features, we introduced a new method for determining \teffns. The heart of this method involves group fitting of hundreds of well-selected spectral features that are not affected by various caveats (continuum mismatch and other inconsistencies between observations and models), as a function of wavelength. Through this method, we determined \teff for different fixed metallicities, and found the optimal \teff with the lowest variation across different wavelengths. We determined \teffns\,=\,3231\,$\pm$\,21\,K for Barnard's star, consistent with the interferometric value of 3238\,$\pm$\,11\,K, which is the most reliable method for \teff determination. Additionally, we showed how previous spectroscopic works might have under- or over-estimated \teffns, possibly due to not considering the various caveats discussed in this work. 

Next, using the determined \teff and fixing \logg from the literature, and utilizing the NIST and BT-Settl PHOENIX line lists, we measured the chemical abundances of 15 different elements. This includes the chemical abundances of Mg, C, Si, and Fe, which play a significant role in exoplanet interior modeling studies. We compared our results with the abundances determined in recent independent literature. Our results were consistent with \cite{ishikawa2022elemental} and partially consistent with \cite{maldonado2020hades}, emphasizing the importance of line selection and methodology in detailed chemical spectroscopy of M dwarfs.

This work emphasizes the need to improve atmosphere models (at least the PHOENIX models used in this work) as there are significant discrepancies between the observations and models. Only a few hundred lines were effectively used in our analysis as a result while several thousands could be used should there be a better agreement between observations and synthetic modes. This detailed and comprehensive analysis should be repeated for other set of models such as MARCS (\citealt{gustafsson2008grid}) and SPHINX (\citealt{iyer2023sphinx}). The high quality NIR spectrum of Barnard's star (broad wavelength coverage, very high SNR and excellent telluric correction) presented here is an ideal data set for such detailed investigations. The full potential of NIR chemical spectroscopy has yet to be harnessed with the development of better atmosphere models.


\acknowledgments

\par We would like to thank the anonymous referee for the constructive comments and thorough review of this manuscript.
\par This work is based on observations obtained at the Canada-France-Hawaii Telescope (CFHT) which is operated from the summit of Maunakea by the National Research Council of Canada, the Institut National des Sciences de l'Univers of the Centre National de la Recherche Scientifique of France, and the University of Hawaii. The observations at the Canada-France-Hawaii Telescope were performed with care and respect from the summit of Maunakea which is a significant cultural and historic site.
\par In this work, we extend our heartfelt gratitude and pay tribute to the late France Allard, whose significant contributions to astrophysics and her instrumental role in the field of synthetic modeling have been invaluable to our research. Throughout this paper, we heavily relied on her pioneering synthetic models, which provided the essential foundation for our analysis and findings. France Allard's groundbreaking work has left a lasting impact on the scientific community, and her legacy continues to inspire current and future generations of researchers.
\par This work is partly funded through the National Science and Engineering Research Council of Canada through the Discovery Grant program and the CREATE training program on New Technologies for Canadian Observatories. We also acknowledge the generous financial support of the Trottier Family Foundation for Trottier Institute for Research on Exoplanets. 
\par We acknowledge funding from the French ANR under contract number ANR18CE310019 (SPlaSH). This work is supported in part by the French National Research Agency in the framework of the Investissements d'Avenir program (ANR-15-IDEX-02), through the funding of the ``Origin of Life'' project of the Grenoble-Alpes University.
\par JFD acknowledges funding from the European Research Council (ERC) under the H2020 research \& innovation program (grant agreement \#740651 NewWorlds) 
\par E.M. acknowledges funding from FAPEMIG under project number APQ-02493-22 and research productivity grant number 309829/2022-4 awarded by the CNPq, Brazil.
\par The research shown here acknowledges use of the Hypatia Catalog Database, an online compilation of stellar abundance data as described in \cite{Hinkel2014}, which was supported by NASA's Nexus for Exoplanet System Science (NExSS) research coordination network and the Vanderbilt Initiative in Data-Intensive Astrophysics (VIDA).

\facilities{CFHT/SPIRou}

\software{\texttt{Astropy} \citep{Astropy_2018}; \texttt{matplotlib} \citep{Hunter_2007}; \texttt{SciPy} \citep{Virtanen_2020}; \texttt{NumPy} \citep{Harris_2020}}.

\pagebreak

\bibliography{BarnardStar}{}
\bibliographystyle{aasjournal}

\appendix
\counterwithin{table}{section}
\counterwithin{figure}{section}

\section{Spectroscopic Data and Abundances of Atomic Lines} \label{sec:appendix}

\begin{figure}[!h]
    \centering\includegraphics[width=0.7\linewidth]{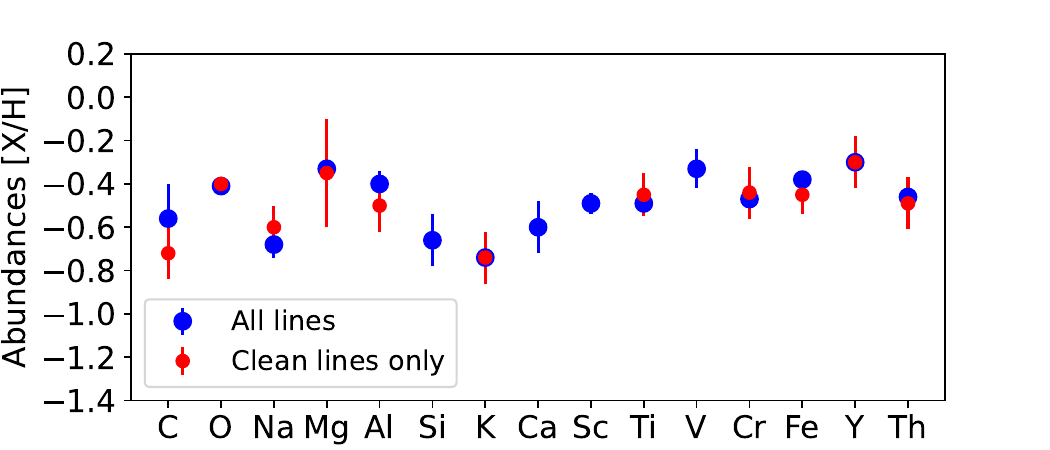}
    \caption{There are no significant abundance differences using either the whole list vs the ones without the potentially contaminated ones and both cases are consistent within their uncertainties.}
    \label{fig:with_without_cont}
\end{figure}

This appendix comprises two tables. Table \ref{table:full_data} presents spectroscopic data (e.g., wavelength and normalized flux) of Barnard's star and a synthetic model with a \teff of 3200\,K, [M/H] of $-$0.5, and \logg of 5 dex. Table \ref{table:refined_linelist} details abundance values for various elemental lines used in this study. 
We compiled this elemental list from NIST database (\citealt{NISTAtomicSpectraDB}) and PHOENIX BT-Settl (\citealt{allard2010model}) line list. In Table \ref{table:refined_linelist}, the column labeled ``Nearby Lines'' identifies the presence of nearby spectral feature based on our line lists. It is important to note that while these nearby lines indicate the proximity of other lines to our target lines, it remains unclear to what extent, if any, these nearby lines influence the strength of the target lines. Due to the inability to individually resolve these lines and their unknown true relative strength, their impact is difficult to quantify. However as a test we have empirically determined that there are no significant abundance differences using either the whole list vs the ones without the potentially contaminated ones (see Figure \ref{fig:with_without_cont}). Also note that the ``Error'' columns represents the standard deviation of all 50 MC measurements for a single line. When the value in this column is zero, it indicates that the noise in the spectrum was too subtle to affect the abundances, suggesting that the systematic uncertainty for the line was below the grid's 0.1 dex metallicity sensitivity.

\begin{deluxetable}{cccc}[ht]
\tablecaption{Spectroscopic Data of Barnard's star and a similar synthetic model\label{table:full_data}}
\tablehead{
\colhead{Wavelength (vacuum, nm)} & \colhead{Data Flux} & \colhead{Error} & \colhead{Model Flux}}
\startdata
970.000 & 0.995 & 0.001 & 0.995\\
970.003 & 0.998 & 0.001 & 0.998\\
970.007 & 1.001 & 0.001 & 1.002\\
970.010 & 0.999 & 0.001 & 1.005\\
970.013 & 1.001 & 0.001 & 1.007\\
970.017 & 0.999 & 0.001 & 1.008\\
970.020 & 0.997 & 0.001 & 1.008\\
970.023 & 0.996 & 0.001 & 1.008\\
970.026 & 0.995 & 0.001 & 1.006\\
970.030 & 0.997 & 0.001 & 1.002\\
\ldots & \ldots & \ldots & \ldots \\
\enddata
\tablecomments{Table~\ref{table:full_data} is published in its entirety in machine-readable format.\\
The synthetic model used for comparison has a \teff of 3200\,K, [M/H] of $-$0.5, and \logg of 5 dex.}
\end{deluxetable}

\begin{table}[b!]
\end{table}

\vspace{5cm}

\begin{longtable}{cccccccc}
\caption{Characteristics of elemental spectral lines\label{table:refined_linelist}} \\
\hline
\colhead{Element} & \colhead{Central Wavelength} & \colhead{[X/H]} & \colhead{Error} & \colhead{Line Depth} & \colhead{Ref} & \colhead{Nearby Lines} \\
\colhead{} & \colhead{(vacuum, nm)} & \colhead{(dex)} & \colhead{(dex)} & \colhead{} & \colhead{} & \colhead{} \\
\hline
\endhead
\hline
\endfoot
     Fe I &   1122.396 &   $-$0.19 &    0.040 &     0.05  &  NIST  & ---\\
     Fe I &   1130.195 &   $-$0.65 &    0.050 &     0.16  &  Ph-BT & ---\\
     Fe I &   1288.329 &   $-$0.30 &    0.020 &     0.16  &  Ph-BT & ---\\
     Fe I &   1400.831 &   $-$0.40 &    0.000 &     0.24  &  NIST  & OH \\
     Fe I &   1467.038 &   $-$0.40 &    0.010 &     0.17  &  NIST  & OH\\
     Fe I &   1538.831 &   $-$0.10 &    0.000 &     0.13  &  NIST  & OH\\
     Fe I &   1543.184 &   $-$0.40 &    0.038 &     0.21  &  NIST  & OH\\
     Fe I &   1578.104 &   $-$0.50 &    0.000 &     0.22  &  NIST  & OH\\
     Fe I &   1590.200 &   $-$0.49 &    0.027 &     0.28  &  NIST  & OH\\
     Fe I &   1591.714 &   $-$0.30 &    0.000 &     0.27  &  NIST  & OH, Mg, Cr\\
     Fe I &   1604.710 &   $-$0.49 &    0.035 &     0.07  &  NIST  & ---\\
     Fe I &   1607.403 &   $-$0.60 &    0.000 &     0.26  &  NIST  & OH\\
     Fe I &   1619.475 &   $-$0.41 &    0.027 &     0.29  &  NIST  & OH\\
     Fe I &   1623.006 &   $-$0.30 &    0.000 &     0.13  &  NIST  & OH\\
     Fe I &   1623.609 &   $-$0.44 &    0.048 &     0.11  &  NIST  & OH\\
     Fe I &   1634.160 &   $-$0.40 &    0.000 &     0.08  &  NIST  & OH\\
     Fe I &   1661.217 &   $-$0.20 &    0.000 &     0.22  &  NIST  & OH\\
     Fe I &   1672.364 &   $-$0.50 &    0.000 &     0.30  &  NIST  & Al, OH\\
     Fe I &   1675.373&   $-$0.40 &    0.000 &     0.20  &  NIST  & OH\\
     Fe I &   1688.363 &   $-$0.40 &    0.000 &     0.20  &  NIST  & OH, Th\\
     Fe I &   1688.941 &   $-$0.30 &    0.000 &     0.18  &  NIST  & OH\\
     Fe I &   1689.980 &   $-$0.40 &    0.000 &     0.23  &  NIST  & OH\\
     Fe I &   1690.350&   $-$0.30 &    0.000 &     0.25  &  NIST  & OH\\
     Fe I &   1705.684 &   $-$0.20 &    0.000 &     0.20  &  NIST  & OH\\
     Fe I &   1733.882 &   $-$0.30 &    0.000 &     0.17  &  NIST  & OH\\
     Fe I &   2179.433 &   $-$0.19 &    0.027 &     0.10  &  NIST  & H$_2$O\\
     Fe I &   2294.137 &   $-$0.70 &    0.000 &     0.20  &  NIST  & CO\\
     Fe I &   2323.967 &   $-$0.60 &    0.000 &     0.17  &  NIST  & ---\\
     Fe I &   2387.064 &   $-$0.30 &    0.000 &     0.29  &  NIST  & H$_2$O\\
     Mg I &   1488.160 &   $-$0.10 &    0.000 &     0.16  &  Ph-BT & ---\\
     Mg I &   1591.718 &   $-$0.30 &    0.000 &     0.27  &  NIST  & OH, Fe, Cr\\
     Mg I &   1711.330 &   $-$0.60 &    0.050 &     0.17  &  Ph-BT & ---\\
     Ti I &   1178.378 &   $-$0.63 &    0.045 &     0.16  &  Ph-BT & ---\\
     Ti I &   1260.371 &   $-$0.70 &    0.000 &     0.14  &  Ph-BT & H$_2$O\\
     Ti I &   1267.458 &   $-$0.58 &    0.039 &     0.15  &  Ph-BT & ---\\
     Ti I &   1274.840 &   $-$0.60 &   0.000  &     0.05  &  Ph-BT & H$_2$O, Cr\\
     Ti I &   1281.499 &   $-$0.33 &    0.056 &     0.12  &  Ph-BT & ---\\
     Ti I &   1292.342 &   $-$0.60 &    0.050 &     0.08  &  Ph-BT & ---\\
     Ti I &   1397.680 &   $-$0.50 &    0.000 &     0.22  &  NIST  & OH\\
     Ti I &   1466.894 &   $-$0.11 &    0.070 &     0.06  &  NIST  & ---\\
     Ti I &   1565.798  &    $-$0.40 &    0.000 &     0.25  &   NIST & OH\\
     Ti I &   1573.491  &    $-$0.59 &    0.032 &     0.27 &    NIST & OH \\
     Ti I &   1606.950  &    $-$0.49 &    0.031 &     0.26 &    NIST & OH\\
     Ti I &   1635.652  &    $-$0.40 &    0.000 &     0.26 &    NIST & OH\\
     Ti I &   1661.012  &    $-$0.40 &    0.000 &     0.19 &    NIST & OH\\
     Ti I &   1993.948  &    $-$0.60 &    0.001 &     0.26 &    NIST & H$_2$O, Ca\\
     Ti I &   2223.892  &    $-$0.50 &    0.000 &     0.18 &    Ph-BT& H$_2$O\\
     Ti I &   2324.308  &    $-$0.60 &    0.000 &     0.16 &    NIST & CO\\
     Ti I &   2328.639  &    $-$0.40 &    0.000 &     0.13 &    NIST & CO, Sc\\ 
     Ti I &   2428.846  &    $-$0.30 &    0.000 &     0.14 &    Ph-BT& H$_2$O\\
     Cr I &   1147.610  &    $-$0.18 &    0.022 &     0.14 &    Ph-BT& V, H$_2$O\\
     Cr I &   1148.770  &    $-$0.44 &    0.048 &     0.17 &    Ph-BT& ---\\
     Cr I &   1161.369  &    $-$0.60 &    0.000 &     0.22 &    Ph-BT& TiO, H$_2$O\\
     Cr I &   1253.620  &    $-$0.57 &    0.022 &     0.07 &    Ph-BT& TiO\\
     Cr I &   1274.844  &    $-$0.60 &    0.000 &     0.05 &    NIST& H$_2$O, Ti\\
     Cr I &   1294.060  &    $-$0.47 &    0.022 &     0.07 &    Ph-BT& TiO\\
     Cr I &   1320.475  &    $-$0.60 &    0.000 &     0.07 &    Ph-BT& H$_2$O\\
     Cr I &   1591.731  &    $-$0.30 &    0.000 &     0.27 &    NIST & OH, Mg, Fe\\
     Na I &   1083.787  &    $-$0.50 &    0.010 &     0.16 &    NIST & ---\\
     Na I &   1268.265  &    $-$0.70 &    0.000 &     0.29 &    NIST & H$_2$O, TiO\\
     Na I &   2206.248  &    $-$0.80 &    0.000 &     0.40 &    NIST & H$_2$O\\
     Na I &   2338.555  &    $-$0.70 &    0.000 &     0.26 &    NIST & ---\\
     Ca I &   1281.957  &    $-$0.60 &    0.000 &     0.17 &    NIST & H$_2$O\\
     Ca I &   1993.914  &    $-$0.60 &    0.001 &     0.26 &    NIST & Ti, H$_2$O\\
     Al I &   1125.628  &    $-$0.30 &    0.022 &     0.22 &    Ph-BT& V, H$_2$O\\
     Al I &   1125.796  &    $-$0.30 &    0.036 &     0.27 &    Ph-BT& Th, TiO \\
     Al I &   1672.353  &    $-$0.50 &    0.000 &     0.30 &    Ph-BT& Fe, OH\\
     Al I &   1675.513  &    $-$0.50 &    0.000 &     0.27 &    Ph-BT& ---\\
     Si I &   1075.231  &    $-$0.66 &    0.010 &     0.05 &    Ph-BT& Th\\
     C I  &   2296.595  &    $-$0.72 &    0.040 &     0.05 &    NIST & ---\\
     C I  &   2354.499  &    $-$0.40 &    0.026 &     0.27 &    NIST & CO\\
     Sc I &   1626.484  &    $-$0.57 &    0.045 &     0.25 &    NIST & OH\\
     Sc I &   1645.469  &    $-$0.50 &    0.014 &     0.25 &    NIST & OH\\
     Sc I &   2328.640  &    $-$0.40 &    0.000 &     0.13 &    NIST & CO, Ti\\ 
     V I  &   1125.610  &    $-$0.30 &    0.022 &     0.22 &    Ph-BT& Al \\
     V I  &   1147.621  &    $-$0.18 &    0.022 &     0.14 &    NIST & Cr\\
     V I  &   2131.777  &    $-$0.50 &    0.043 &     0.06 &    NIST & H$_2$O\\
     K I  &   1102.290  &    $-$0.74 &    0.040 &     0.15 &    NIST & ---\\
     Y I  &   2399.700  &    $-$0.30 &    0.010 &     0.21 &    Ph-BT& ---\\
     Th I &   1075.237  &    $-$0.66 &    0.010 &     0.05 &    Ph-BT& Si\\
     Th I &   1125.816  &    $-$0.30 &    0.036 &     0.27 &    NIST & Al, TiO \\
     Th I &   1391.583  &    $-$0.70 &    0.017 &     0.09 &    NIST & OH\\
     Th I &   1391.899  &    $-$0.35 &    0.050 &     0.20 &    NIST & OH\\
     Th I &   1466.534 &    $-$0.50 &    0.000 &     0.28 &    NIST & OH\\
     Th I &   1557.637  &    $-$0.50 &    0.000 &     0.29 &    NIST & OH\\
     Th I&   1635.060 &     $-$0.27 &    0.048 &     0.17 &    NIST & OH\\
     Th I&   1652.792 &     $-$0.50 &    0.000 &     0.23 &    NIST & OH\\
     Th I&   1673.448 &     $-$0.37 &    0.045 &     0.16 &    NIST & OH\\
     Th I &   1687.645 &    $-$0.40 &    0.000 &     0.21 &    NIST & OH\\
     Th I &   1688.391 &    $-$0.40 &    0.000 &     0.20 &    NIST & OH, Fe\\
     Th I &   1700.934 &    $-$0.60 &    0.022 &     0.09 &    NIST & OH\\
     Th I &   1710.822 &    $-$0.49 &    0.058 &     0.08 &    NIST & ---\\
       OH &   1391.577 &    $-$0.70 &    0.017 &     0.09 &   Ph-BT & Th\\       
       OH &   1391.888 &    $-$0.35 &    0.050 &     0.20 &   Ph-BT & Th\\
       OH &   1391.953 &    $-$0.40 &    0.000 &     0.21 &   Ph-BT & ---\\
       OH &   1393.282 &    $-$0.38 &    0.034 &     0.22 &   Ph-BT & ---\\
       OH &   1394.495 &    $-$0.40 &    0.000 &     0.17 &   Ph-BT & ---\\
       OH &   1397.657 &    $-$0.50 &    0.000 &     0.22 &   Ph-BT & Ti\\
       OH &   1400.822 &    $-$0.40 &    0.000 &     0.24 &   Ph-BT & Fe\\
       OH &   1405.901 &    $-$0.45 &    0.050 &     0.23 &   Ph-BT & ---\\
       OH &   1408.694 &    $-$0.60 &    0.000 &     0.19 &   Ph-BT & ---\\
       OH &   1413.419 &    $-$0.70 &    0.020 &     0.15 &   Ph-BT & ---\\
       OH &   1416.298 &    $-$0.42 &    0.040 &     0.20 &   Ph-BT & ---\\
       OH &   1418.596 &    $-$0.50 &    0.000 &     0.16 &   Ph-BT & ---\\
       OH &   1434.455 &    $-$0.50 &    0.051 &     0.24 &   Ph-BT & ---\\
       OH &   1456.402 &    $-$0.44 &    0.048 &     0.26 &   Ph-BT & ---\\
       OH &   1461.844 &    $-$0.37 &    0.044 &     0.15 &   Ph-BT & ---\\
       OH &   1466.515 &    $-$0.50 &    0.000 &     0.28 &   Ph-BT & Th\\
       OH &   1467.031 &    $-$0.40 &    0.010 &     0.17 &   Ph-BT & Fe\\
       OH &   1469.887 &    $-$0.50 &    0.000 &     0.21 &   Ph-BT & ---\\
       OH &   1476.043 &    $-$0.45 &    0.050 &     0.06 &   Ph-BT & ---\\
       OH &   1477.216 &    $-$0.62 &    0.038 &     0.21 &   Ph-BT & ---\\
       OH &   1483.328 &    $-$0.30 &    0.000 &     0.15 &   Ph-BT & ---\\
       OH &   1500.722 &    $-$0.50 &    0.017 &     0.26 &   Ph-BT & ---\\
       OH &   1505.320 &     0.10   &    0.020 &     0.12 &   Ph-BT & ---\\
       OH &   1506.937 &    $-$0.10 &    0.049 &     0.17 &   Ph-BT & ---\\
       OH &   1513.380 &    $-$0.30 &    0.000 &     0.32 &   Ph-BT & ---\\
       OH &   1513.506 &    $-$0.52 &    0.038 &     0.28 &   Ph-BT & ---\\
       OH &   1524.110 &    $-$0.45 &    0.050 &     0.15 &   Ph-BT & ---\\
       OH &   1528.269 &    $-$0.50 &    0.000 &     0.29 &   Ph-BT & ---\\
       OH &   1528.524 &    $-$0.50 &    0.000 &     0.29 &   Ph-BT & ---\\
       OH &   1533.212 &    $-$0.30 &    0.020 &     0.18 &   Ph-BT & ---\\
       OH &   1539.525 &    $-$0.41 &    0.026 &     0.21 &      Ph-BT & ---\\
       OH &   1539.541 &    $-$0.40 &    0.032 &     0.21 &      Ph-BT & ---\\
       OH &   1540.733 &    $-$0.24 &    0.050 &     0.15 &      Ph-BT & ---\\
       OH &   1542.985 &    $-$0.30 &    0.000 &     0.12 &      Ph-BT & ---\\
       OH &   1543.171 &    $-$0.58 &    0.038 &     0.21 &      Ph-BT & Fe\\
       OH &   1546.814 &    $-$0.17 &    0.045 &     0.10 &      Ph-BT & ---\\
       OH &   1550.999 &    $-$0.50 &    0.000 &     0.21 &      Ph-BT & ---\\
       OH &   1557.023 &    $-$0.30 &    0.010 &     0.14 &      Ph-BT & ---\\
       OH &   1557.633 &    $-$0.50 &    0.000 &     0.29 &      Ph-BT & Th\\
       OH &   1559.783 &    $-$0.17 &    0.045 &     0.08 &      Ph-BT & ---\\
       OH &   1563.095 &    $-$0.32 &    0.038 &     0.28 &      Ph-BT & ---\\
       OH &   1563.168 &    $-$0.30 &    0.014 &     0.25 &      Ph-BT & ---\\
       OH &   1565.778 &    $-$0.40 &    0.000 &     0.25 &      Ph-BT & Ti\\
       OH &   1573.474 &    $-$0.59 &    0.032 &     0.27 &      Ph-BT & Ti\\       
       OH &   1578.115 &    $-$0.50 &    0.000 &     0.22 &      Ph-BT & Fe\\
       OH &   1583.314 &    $-$0.30 &    0.014 &     0.13 &      Ph-BT & ---\\
       OH &   1583.335 &    $-$0.30 &    0.010 &     0.13 &      Ph-BT & ---\\
       OH &   1590.206 &    $-$0.49 &    0.027 &     0.28 &      Ph-BT & Fe\\
       OH &   1591.708 &    $-$0.30 &    0.000 &     0.27 &      Ph-BT  & Mg, Fe, Cr\\
       OH &   1604.126 &    $-$0.50 &    0.010 &     0.24 &      Ph-BT & ---\\
       OH &   1604.292 &    $-$0.50 &    0.000 &     0.24 &      Ph-BT & ---\\
       OH &   1606.943 &    $-$0.49 &    0.031 &     0.26 &      Ph-BT & Ti\\
       OH &   1607.394 &    $-$0.60 &    0.000 &     0.26 &      Ph-BT & Fe\\
       OH &   1607.984 &    $-$0.40 &    0.017 &     0.15 &      Ph-BT & ---\\
       OH &   1612.829 &    $-$0.35 &    0.050 &     0.12 &      Ph-BT & ---\\
       OH &   1619.454 &    $-$0.41 &    0.027 &     0.29 &      Ph-BT & Fe\\
       OH &   1621.162 &    $-$0.43 &    0.043 &     0.25 &      Ph-BT & ---\\
       OH &   1622.996 &    $-$0.30 &    0.000 &     0.13 &      Ph-BT & ---\\
       OH &   1623.489 &    $-$0.37 &    0.045 &     0.13 &       Ph-BT & ---\\
       OH &   1623.587 &    $-$0.44 &    0.048 &     0.11 &       Ph-BT & Fe\\
       OH &   1626.459 &    $-$0.57 &    0.045 &     0.25 &       Ph-BT & Sc\\       
       OH &   1627.024 &    $-$0.55 &    0.050 &     0.11 &       Ph-BT & ---\\
       OH &   1631.692 &    $-$0.47 &    0.044 &     0.13 &       Ph-BT & ---\\
       OH &   1631.736 &    $-$0.49 &    0.045 &     0.13 &       Ph-BT & ---\\
       OH &   1634.149 &    $-$0.40 &    0.000 &     0.08 &       Ph-BT & Fe\\
       OH &   1635.065 &    $-$0.27 &    0.048 &     0.17 &       Ph-BT & Th\\
       OH &   1635.196 &    $-$0.51 &    0.034 &     0.13 &       Ph-BT & ---\\
       OH &   1635.670 &    $-$0.40 &    0.000 &     0.26 &       Ph-BT & Ti\\
       OH &   1636.904 &    $-$0.44 &    0.049 &     0.26 &       Ph-BT & ---\\
       OH &   1645.257 &    $-$0.50 &    0.000 &     0.23 &       Ph-BT & ---\\
       OH &   1645.488 &    $-$0.50 &    0.014 &     0.25 &       Ph-BT & Sc\\
       OH &   1646.053 &    $-$0.50 &    0.000 &     0.24 &       Ph-BT & ---\\
       OH &   1647.734 &    $-$0.40 &    0.000 &     0.15 &       Ph-BT & ---\\
       OH &   1652.804 &    $-$0.50 &    0.000 &     0.23 &       Ph-BT & Th\\
       OH &   1653.074 &    $-$0.51 &    0.027 &     0.24 &       Ph-BT & ---\\
       OH &   1653.913 &    $-$0.50 &    0.000 &     0.25 &       Ph-BT & ---\\
       OH &   1658.581 &    $-$0.24 &    0.022 &     0.17 &       Ph-BT & ---\\
       OH &   1658.686 &    $-$0.30 &    0.010 &     0.17 &       Ph-BT & ---\\
       OH &   1661.000 &    $-$0.40 &    0.000 &     0.19 &       Ph-BT & Ti\\       
       OH &   1661.205 &    $-$0.20 &    0.000 &     0.22 &       Ph-BT & Fe\\
       OH &   1665.922 &    $-$0.55 &    0.050 &     0.22 &       Ph-BT & ---\\
       OH &   1666.055 &    $-$0.60 &    0.014 &     0.23 &       Ph-BT & ---\\
       OH &   1666.672 &    $-$0.50 &    0.000 &     0.22 &       Ph-BT & ---\\
       OH &   1670.892 &    $-$0.50 &    0.000 &     0.22 &       Ph-BT & ---\\
       OH &   1671.895 &    $-$0.52 &    0.041 &     0.23 &       Ph-BT & ---\\
       OH &   1672.342 &    $-$0.50 &    0.000 &     0.30 &       Ph-BT & Al, Fe\\
       OH &   1673.435 &    $-$0.37 &    0.045 &     0.16 &       Ph-BT & Th\\       
       OH &   1675.385 &    $-$0.40 &    0.000 &     0.20 &       Ph-BT & Fe\\       
       OH &   1675.630 &    $-$0.60 &    0.010 &     0.19 &       Ph-BT & ---\\
       OH &   1687.651 &    $-$0.40 &    0.000 &     0.21 &       Ph-BT & Th\\       
       OH &   1687.691 &    $-$0.50 &    0.010 &     0.22 &       Ph-BT & ---\\
       OH &   1688.372 &    $-$0.40 &    0.000 &     0.20 &       Ph-BT & Fe, Th\\
       OH &   1688.913 &    $-$0.30 &    0.000 &     0.18 &      Ph-BT & Fe\\
       OH &   1689.087 &    $-$0.40 &    0.000 &     0.18 &      Ph-BT & ---\\
       OH &   1689.978 &    $-$0.40 &    0.000 &     0.23 &      Ph-BT & Fe\\       
       OH &   1690.339 &    $-$0.30 &    0.000 &     0.25 &      Ph-BT & Fe\\       
       OH &   1690.891 &    $-$0.40 &    0.000 &     0.23 &      Ph-BT & ---\\
       OH &   1691.027 &    $-$0.40 &    0.000 &     0.18 &      Ph-BT & ---\\
       OH &   1691.388 &    $-$0.50 &    0.014 &     0.23 &      Ph-BT & ---\\
       OH &   1695.517 &    $-$0.20 &    0.000 &     0.10 &      Ph-BT & ---\\
       OH &   1700.910 &    $-$0.60 &    0.022 &     0.09 &      Ph-BT & Th\\       
       OH &   1705.688 &    $-$0.20 &    0.000 &     0.20 &      Ph-BT & Fe\\
       OH &   1707.077 &    $-$0.40 &    0.000 &     0.17 &      Ph-BT & ---\\
       OH &   1710.109 &    $-$0.40 &    0.017 &     0.19 &      Ph-BT & ---\\
       OH &   1710.440 &    $-$0.60 &    0.010 &     0.18 &      Ph-BT & ---\\
       OH &   1710.942 &    $-$0.30 &    0.017 &     0.23 &      Ph-BT & ---\\
       OH &   1711.181 &    $-$0.40 &    0.000 &     0.18 &      Ph-BT & ---\\
       OH &   1717.982 &    $-$0.12 &    0.044 &     0.11 &      Ph-BT & ---\\
       OH &   1732.697 &    $-$0.20 &    0.000 &     0.20 &      Ph-BT & ---\\
       OH &   1733.871 &    $-$0.30 &    0.000 &     0.17 &      Ph-BT & Fe\\
       OH &   1734.501 &    $-$0.66 &    0.048 &     0.17 &      Ph-BT & ---\\
       OH &   1735.167 &    $-$0.26 &    0.049 &     0.22 &      Ph-BT & ---\\
       OH &   1741.871 &    $-$0.39 &    0.037 &     0.16 &      Ph-BT & ---\\
       OH &   1762.373 &    $-$0.30 &    0.000 &     0.15 &      Ph-BT & ---\\
       OH &   1777.190 &    $-$0.10 &    0.000 &     0.15 &      Ph-BT & ---\\
       OH &   1781.951 &    $-$0.10 &    0.000 &     0.17 &      Ph-BT & ---\\
       OH &   1783.556 &    $-$0.41 &    0.022 &     0.10 &      Ph-BT & ---\\   
       OH &   1828.562 &    $-$0.05 &    0.050 &     0.24 &      Ph-BT & ---\\
\end{longtable}



\clearpage

\section{Full Spectrum Comparison} \label{sec:appendix2}

Full wavelength comparison of Barnard's star spectrum against a synthetic model with a \teff of 3200\,K, [M/H] of $-$0.5, and \logg of 5 dex.

\begin{figure}[!h]
    \centering
    \rotatebox{90}{\includegraphics[width=1.1\linewidth]{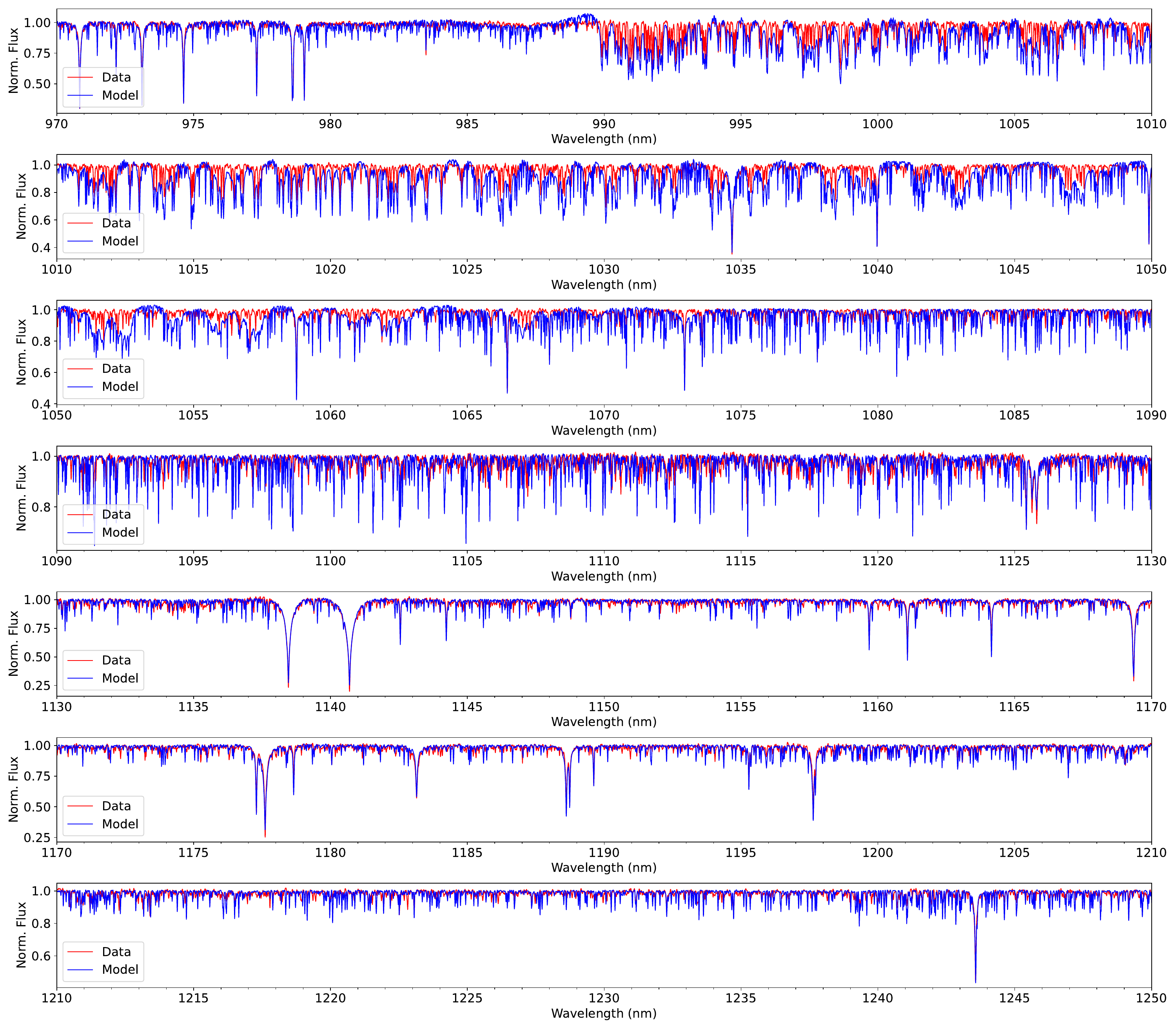}}
    \caption{Comparison between the observed data (red) and the model (blue) with \teffns, \logg, and metallicity of 3200\,K, 5.0 dex and $-$0.5 dex, respectively. The continuum mismatch (see Section \ref{Cont}) is evident within 985-1068 nm.}
    \label{fig:full_comp1}
\end{figure}

\begin{figure}[!h]
    \centering
    \rotatebox{90}{\includegraphics[width=1.1\linewidth]{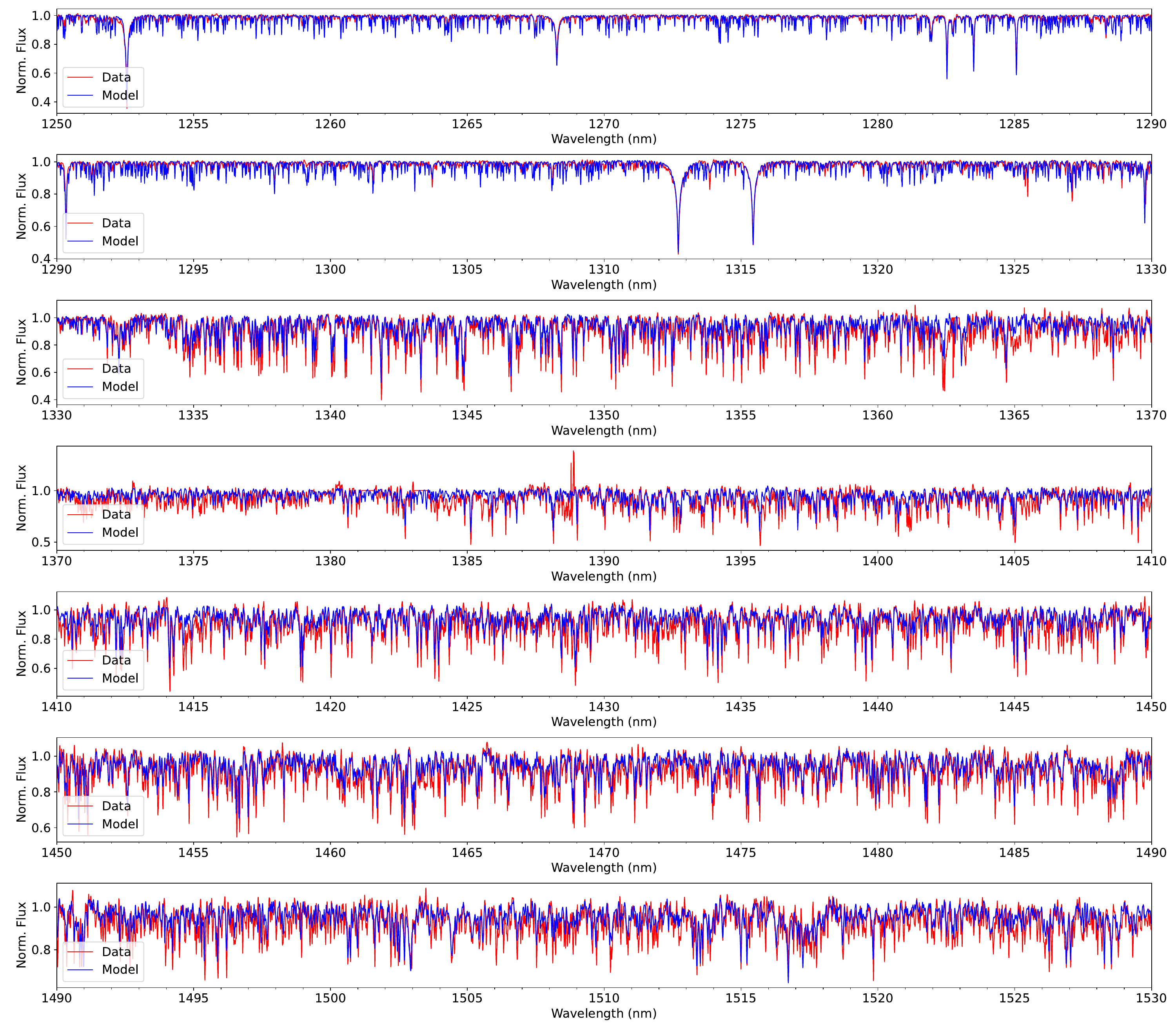}}
    \caption{Comparison between the observed data (red) and the model (blue) with \teffns, \logg, and metallicity of 3200\,K, 5.0 dex and $-$0.5 dex, respectively.}
    \label{fig:full_comp2}
\end{figure}

\begin{figure}[!h]
    \centering
    \rotatebox{90}{\includegraphics[width=1.1\linewidth]{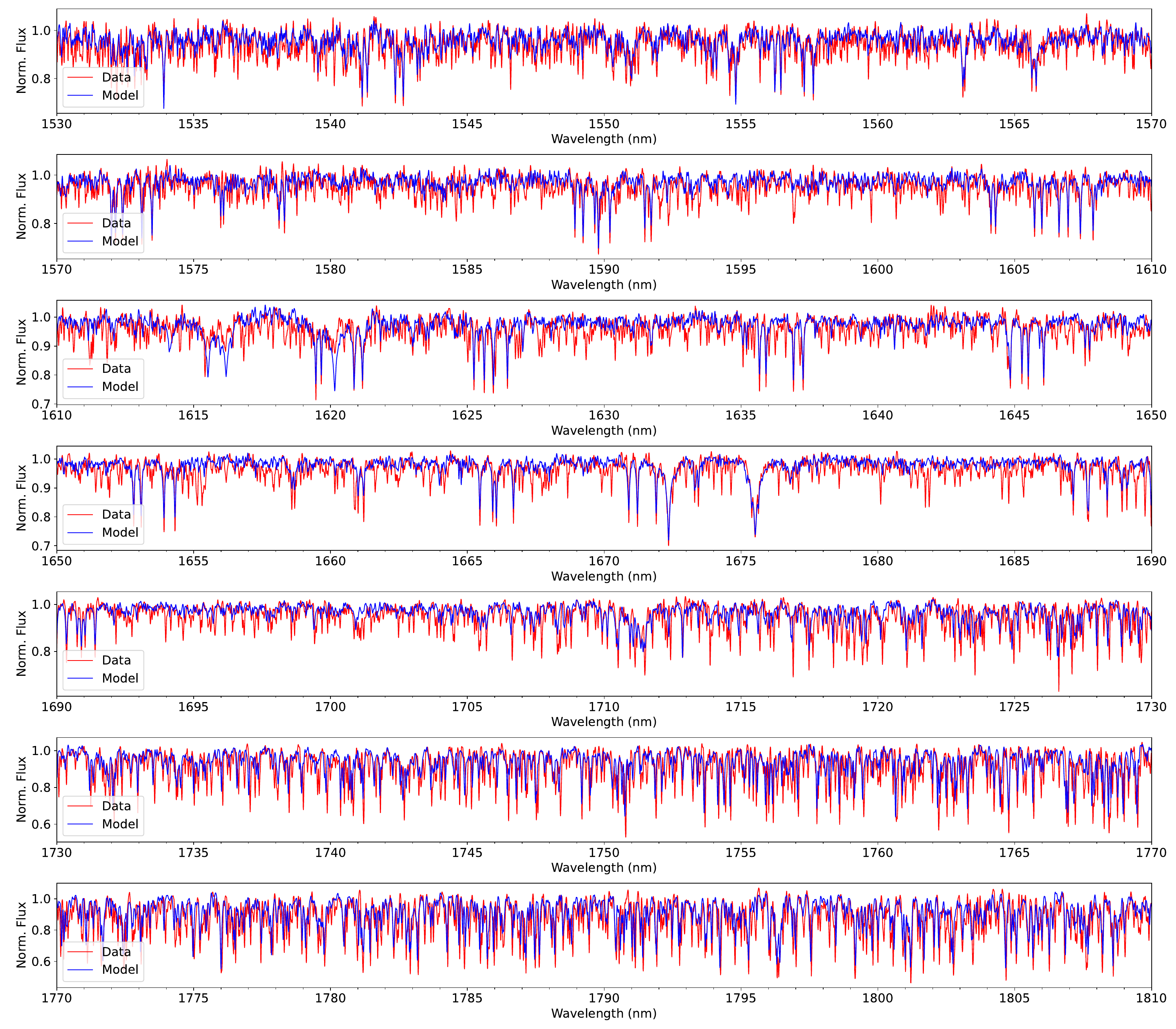}}
    \caption{Comparison between the observed data (red) and the model (blue) with \teffns, \logg, and metallicity of 3200\,K, 5.0 dex and $-$0.5 dex, respectively.}
    \label{fig:full_comp3}
\end{figure}

\begin{figure}[!h]
    \centering
    \rotatebox{90}{\includegraphics[width=1.1\linewidth]{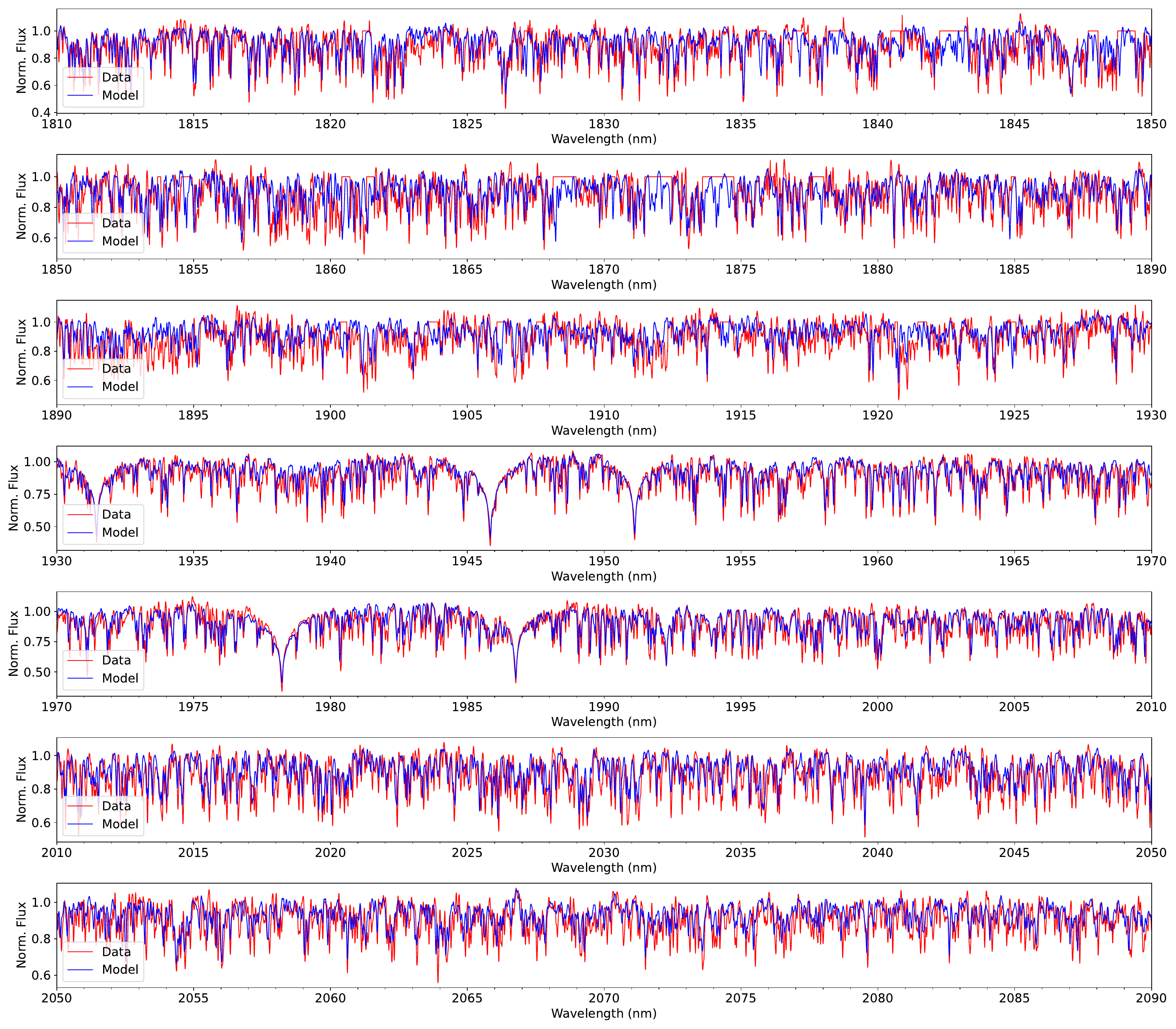}}
    \caption{Comparison between the observed data (red) and the model (blue) with \teffns, \logg, and metallicity of 3200\,K, 5.0 dex and $-$0.5 dex, respectively. The missing data in locations such as 1840 nm are due to the telluric corrections.}
    \label{fig:full_comp4}
\end{figure}

\begin{figure}[!h]
    \centering
    \rotatebox{90}{\includegraphics[width=1.1\linewidth]{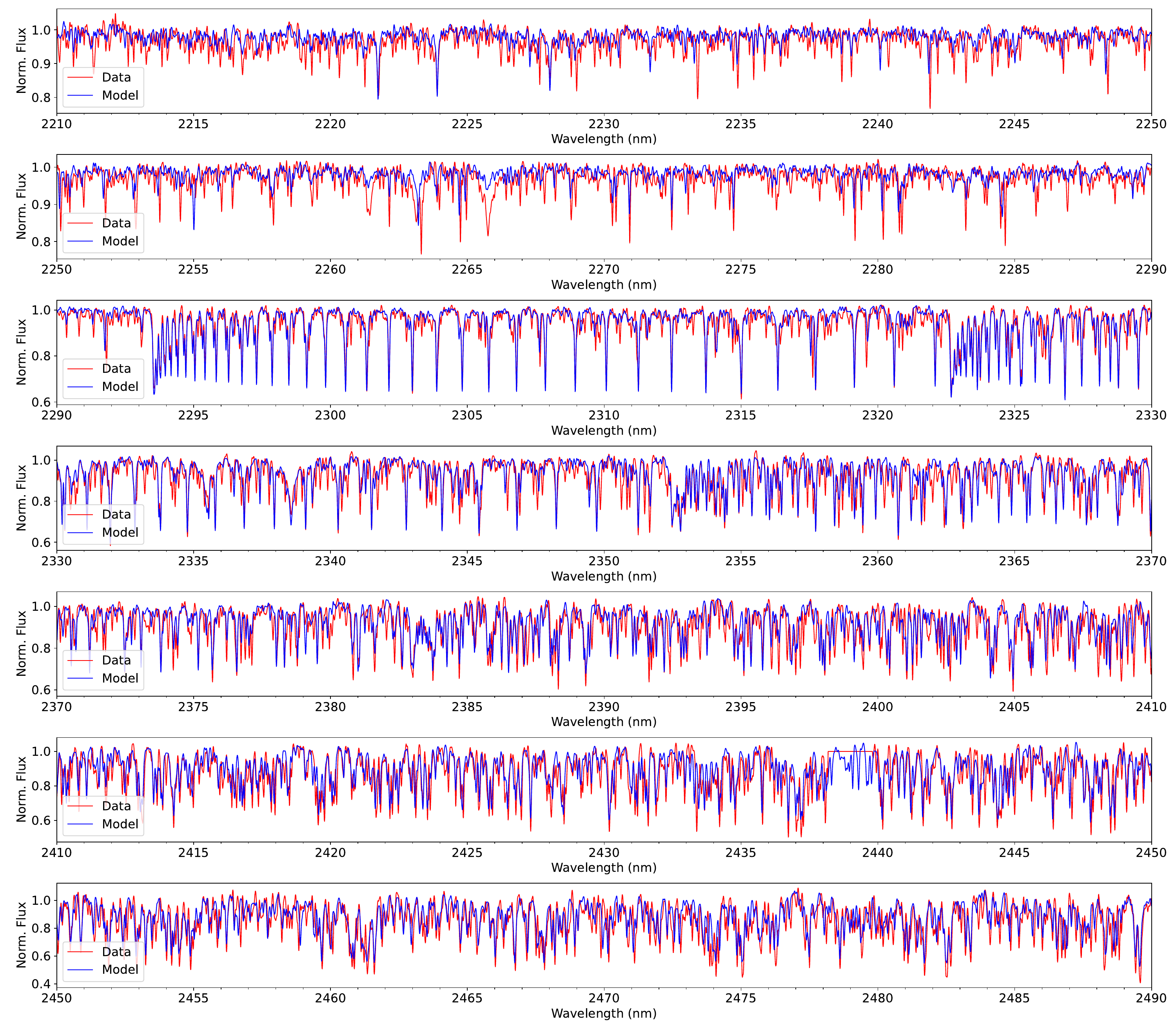}}
    \caption{Comparison between the observed data (red) and the model (blue) with \teffns, \logg, and metallicity of 3200\,K, 5.0 dex and $-$0.5 dex, respectively.}
    \label{fig:full_comp5}
\end{figure}

\end{document}